\documentclass{aa}  

\newcommand{\ihor}{$\iota$\,Hor}
\usepackage{graphicx}
\usepackage{txfonts}
\begin{document}

   \title{Multi-wavelength variability of the young solar analog $\iota$\,Hor}
     
   \subtitle{X-ray cycle, star spots, flares, and UV emission}
   
   \author{J. Sanz-Forcada\inst{1}
          \and
          B. Stelzer\inst{2,3} \and M. Coffaro\inst{2} \and
          S. Raetz\inst{2} \and J. D. Alvarado-G\'omez\inst{4}
          }

   \institute{Departamento de Astrof\'{i}sica,
     Centro de Astrobiolog\'{i}a (CSIC-INTA), ESAC Campus, Camino bajo
     del Castillo s/n, 
     E-28692 Villanueva de la Ca\~nada, Madrid, Spain; \\
     \email{jsanz@cab.inta-csic.es}
     \and
     Institut f\"ur Astronomie and Astrophysik T\"ubingen (IAAT), Eberhard-Karls Universit\"at T\"ubingen, Sand 1, D-72076, Germany;
     \email{stelzer@astro.uni-tuebingen.de,coffaro@astro.uni-tuebingen.de,raetz@astro.uni-tuebingen.de}
     \and
     INAF -- Osservatorio Astronomico di Palermo
     G. S. Vaiana, Piazza del Parlamento 1, Palermo, I-90134 Italy;
     \and
     Center for Astrophysics $|$ Harvard \& Smithsonian, 60 Garden
     Street, Cambridge, MA 02138, USA
     \email{jalvarad@cfa.harvard.edu}}
   \date{Received ; accepted }

 
  \abstract
      {Chromospheric activity cycles are common in late-type
        stars; however, only a handful of coronal activity cycles have been
        discovered. \ihor\ is the most active and youngest
        star with known coronal cycles. It is also a young solar
        analog, and we are likely facing the earliest cycles in the
        evolution of solar-like stars, at an age ($\sim600$~Myr) when
        life appeared on Earth.}
   {Our aim is to confirm the $\sim$1.6~yr coronal cycle and
     characterize its stability over time. We use X-ray
     observations of \ihor\ to study the corona of a star
     representing the solar past through
     variability, thermal structure, and coronal abundances.}
   {We analyzed multi-wavelength observations of \ihor\ using {\it
       XMM-Newton}, {\it TESS}, and {\it HST} data.
     We monitored \ihor\ throughout almost seven years in X-rays and in
     two UV bands.
     The summed RGS and STIS
     spectra were used for a detailed thermal structure model, 
     and the determination of coronal abundances.
     We studied rotation and flares in the {\it TESS} light curve.}
   {We find a stable coronal cycle along four complete 
     periods, more than covered in the Sun.
     There is no evidence for a second longer X-ray cycle.
     Coronal abundances are  
     consistent with photospheric values, discarding any effects
     related to the first ionization potential. From the {\it TESS}
     light curve we derived the first photometric measurement of the
     rotation period (8.2~d).
     No flares were detected in the {\it TESS} light curve of \ihor.
     We estimate the probability of having detected zero flares with
     {\it TESS} to be $\sim 2$\%.}
   {We corroborate the presence of an activity cycle of $\sim$1.6~yr in
     \ihor\ in X-rays, more regular than its \ion{Ca}{ii} H\&K
     counterpart. A decoupling of the activity between the northern and
     southern hemispheres of the star might explain the disagreement. The
     inclination of the system would result in an irregular 
     behavior in the chromospheric indicators. The more extended
     coronal material would be less sensitive to this effect.}


   \maketitle
%


\section{Introduction}\label{sect:intro}

Stellar activity is common among late-type stars, due to a combination of
rotation and magnetism. Among the observational consequences are
  stellar spots, protuberances, coronal loops, flares, and
activity cycles. The photospheric spots allow us to measure stellar
rotation. Rotation slows down with time, thus the rotational period
  is a link with the stellar age \citep{sku72}. Stellar 
flares can reveal information of the dimensions on the active regions
on the surface of the stars
\citep[][and references therein]{san07,rea14}, but our
knowledge of spot coverage and distribution over the stellar surface
is still limited. The present-day Sun shows activity cycles with
  a dominating periodicity lasting
$\sim$11 yr, although with some range in its duration 
\citep[e.g.,][]{hat10} and well-known irregularities, such as the
Maunder minimum, which also indicates a clear connection between the
solar activity and the Earth's climate, evident also in the infrared
energy budget of the Earth's
thermosphere \citep[][and references therein]{fri91,mly18}.
It has been possible for some
decades to search for the activity cycles in other late-type
stars, allowing us to explore a range of values in the general
variables behind the existence of these cycles. The Mount Wilson
\ion{Ca}{ii} H\&K S-index survey was used by \citet[][and references
  therein]{bal95} to find that $\sim$60\% of the 
main-sequence stars with spectral types from F to early M show chromospheric
cycles with periods in the range $2.5-25$\,yr. The very active (also
youngest) stars tend to show irregular chromospheric variability 
rather than cycles. More mature stars with moderate levels of
activity tend to show more stable cycles, while old inactive stars
have no indication of cycles, which might be interpreted as a stage
similar to the Maunder minimum \citep{tes15}.

%
\begin{table*}
\caption[]{XMM-Newton observation log, OM flux density, exposure times 
  of the EPIC and RGS instruments, and results of EPIC spectral
  fits}\label{tab:xraylog}  
\tabcolsep 2.5 pt
\renewcommand{\arraystretch}{1.3}  
\begin{center}
\begin{small}
\begin{tabular}{ccccccccccc}
\hline \hline
Date & XMM & \multicolumn{2}{c}{OM Flux ($\times
  10^{-12}$~erg~s$^{-1}$~cm$^{-2}$ \AA$^{-1}$)} &
\multicolumn{4}{c}{$t_{\rm exp}$ (ks)} & $\log T_{1,2}$ & $\log EM_{1,2}$ & $L_{\rm X}$ \\
     & Rev. & UVM2 & UVW2 & pn & MOS 1 & MOS 2 & RGS & (K) & (cm$^{-3}$) & ($\times 10^{28}$~erg s$^{-1}$) \\
\hline
2011-05-16 & 2094 & 1.899$\pm$0.040 & \dots & 5.0 & 7.32 & 7.2 & 15.6 & 6.57$^{+0.03}_{-0.06}$, 6.88$^{+0.03}_{-0.05}$ & 51.18$^{+0.05}_{-0.12}$, 50.79$^{+0.19}_{-0.13}$ & 5.65$\pm$0.07 \\
2011-06-11 & 2107 & \dots & \dots & 5.9 & 14.2 & 14.3 & 29.4 & 6.58$^{+0.02}_{-0.02}$, 6.92$^{+0.02}_{-0.02}$ & 51.26$^{+0.03}_{-0.03}$, 50.90$^{+0.05}_{-0.05}$ & 7.08$\pm$0.06 \\
2011-07-09 & 2121 & 1.960$\pm$0.040 & \dots & 6.9 & 9.18 & 9.2 & 19.2 & 6.58$^{+0.02}_{-0.04}$, 6.89$^{+0.01}_{-0.05}$ & 51.21$^{+0.03}_{-0.10}$, 50.78$^{+0.17}_{-0.07}$ & 5.93$\pm$0.06 \\
2011-08-04 & 2134 & 1.914$\pm$0.038 & \dots & 6.1 & 8.47 & 8.41 & 17.6 & 6.59$^{+0.03}_{-0.02}$, 6.95$^{+0.03}_{-0.02}$ & 51.27$^{+0.04}_{-0.04}$, 50.96$^{+0.05}_{-0.10}$ & 7.47$\pm$0.07 \\
2011-11-20 & 2188 & 1.928$\pm$0.048 & \dots & 5.5 & 7.55 & 7.67 & 16.0 & 6.59$^{+0.03}_{-0.02}$, 6.96$^{+0.03}_{-0.02}$ & 51.19$^{+0.05}_{-0.05}$, 51.02$^{+0.05}_{-0.08}$ & 7.24$\pm$0.07 \\
2011-12-18 & 2202 & 1.877$\pm$0.045 & \dots & 4.86 & 6.95 & 7.03 & 14.6 & 6.57$^{+0.02}_{-0.03}$, 6.94$^{+0.03}_{-0.02}$ & 51.17$^{+0.04}_{-0.04}$, 50.78$^{+0.07}_{-0.09}$ & 5.52$\pm$0.07 \\
2012-01-15 & 2216 & 1.822$\pm$0.042 & \dots & 4.56 & 6.64 & 6.68 & 13.9 & 6.59$^{+0.04}_{-0.07}$, 6.88$^{+0.04}_{-0.05}$ & 51.05$^{+0.07}_{-0.16}$, 50.75$^{+0.20}_{-0.18}$ & 4.65$\pm$0.06 \\
2012-02-10 & 2229 & 1.835$\pm$0.034 & \dots & 7.42 & 9.72 & 9.91 & 20.4 & 6.53$^{+0.05}_{-0.07}$, 6.83$^{+0.07}_{-0.05}$ & 50.99$^{+0.09}_{-0.13}$, 50.69$^{+0.17}_{-0.26}$ & 3.86$\pm$0.05 \\
2012-05-19 & 2279 & 1.844$\pm$0.043 & \dots & 6.19 & 8.48 & 8.51 & 17.6 & 6.48$^{+0.06}_{-0.05}$, 6.85$^{+0.05}_{-0.02}$ & 50.98$^{+0.09}_{-0.07}$, 50.81$^{+0.07}_{-0.17}$ & 4.34$\pm$0.06 \\
2012-06-29 & 2299 & 1.790$\pm$0.029 & \dots & 9.73 & 12.3 & 12.3 & 25.6 & 6.55$^{+0.02}_{-0.12}$, 6.90$^{+0.02}_{-0.07}$ & 51.11$^{+0.03}_{-0.07}$, 50.77$^{+0.23}_{-0.06}$ & 5.01$\pm$0.05 \\
2012-08-09 & 2320 & 1.853$\pm$0.036 & \dots & 6.7 & 8.73 & 9.0 & 18.6 & 6.48$^{+0.07}_{-0.05}$, 6.83$^{+0.04}_{-0.02}$ & 50.95$^{+0.11}_{-0.07}$, 50.93$^{+0.06}_{-0.15}$ & 4.83$\pm$0.06 \\
2012-11-18 & 2371 & 1.853$\pm$0.031 & \dots & 3.99 & 6.13 & 5.79 & 17.0 & 6.56$^{+0.03}_{-0.10}$, 6.88$^{+0.04}_{-0.06}$ & 51.13$^{+0.06}_{-0.17}$, 50.78$^{+0.22}_{-0.14}$ & 5.24$\pm$0.07 \\
2012-12-20 & 2387 & 1.851$\pm$0.044 & \dots & 4.49 & 5.53 & 6.46 & 13.6 & 6.57$^{+0.03}_{-0.08}$, 6.88$^{+0.03}_{-0.05}$ & 51.08$^{+0.07}_{-0.16}$, 50.83$^{+0.17}_{-0.14}$ & 5.17$\pm$0.07 \\
2013-02-03 & 2409 & 1.848$\pm$0.036 & \dots & 6.44 & 7.51 & 8.75 & 18.1 & 6.53$^{+0.05}_{-0.09}$, 6.87$^{+0.04}_{-0.04}$ & 51.01$^{+0.09}_{-0.14}$, 50.85$^{+0.13}_{-0.15}$ & 4.69$\pm$0.06 \\
2013-05-20 & 2462 & \dots & 1.673$\pm$0.009 & 4.5 & 6.25 & 6.5 & 13.6 & 6.57$^{+0.03}_{-0.03}$, 6.91$^{+0.02}_{-0.02}$ & 51.16$^{+0.05}_{-0.06}$, 51.00$^{+0.07}_{-0.06}$ & 6.79$\pm$0.08 \\
2013-08-09 & 2503 & \dots & 1.657$\pm$0.006 & 9.86 & 11.8 & 12.3 & 24.8 & 6.58$^{+0.03}_{-0.04}$, 6.87$^{+0.04}_{-0.04}$ & 51.20$^{+0.06}_{-0.09}$, 50.87$^{+0.14}_{-0.16}$ & 6.33$\pm$0.06 \\
2014-02-05 & 2593 & \dots & 1.669$\pm$0.008 & 6.3 & 8.24 & 8.4 & 17.6 & 6.57$^{+0.02}_{-0.04}$, 6.90$^{+0.02}_{-0.05}$ & 51.15$^{+0.03}_{-0.08}$, 50.70$^{+0.16}_{-0.06}$ & 5.07$\pm$0.06 \\
2014-05-18 & 2644 & \dots & 1.677$\pm$0.006 & 9.22 & 11.1 & 11.7 & 24.2 & 6.58$^{+0.01}_{-0.02}$, 6.90$^{+0.02}_{-0.05}$ & 51.13$^{+0.02}_{-0.05}$, 50.50$^{+0.16}_{-0.08}$ & 4.31$\pm$0.05 \\
2014-08-11 & 2687 & \dots & 1.673$\pm$0.007 & 9.31 & 11.1 & 11.8 & 24.4 & 6.60$^{+0.01}_{-0.01}$, 6.91$^{+0.03}_{-0.03}$ & 51.29$^{+0.02}_{-0.03}$, 50.71$^{+0.10}_{-0.09}$ & 6.56$\pm$0.06 \\
2014-11-20 & 2738 & \dots & 1.658$\pm$0.009 & 5.6 & 7.49 & 7.73 & 16.2 & 6.57$^{+0.02}_{-0.02}$, 6.92$^{+0.02}_{-0.02}$ & 51.26$^{+0.03}_{-0.04}$, 51.00$^{+0.05}_{-0.05}$ & 7.69$\pm$0.08 \\
2015-02-06 & 2777 & \dots & 1.683$\pm$0.002 & 4.48 & 6.08 & 6.44 & 13.6 & 6.56$^{+0.04}_{-0.08}$, 6.87$^{+0.03}_{-0.04}$ & 51.18$^{+0.08}_{-0.15}$, 51.04$^{+0.14}_{-0.12}$ & 7.21$\pm$0.08 \\
2015-05-21 & 2829 & 1.819$\pm$0.001 & 1.669$\pm$0.002 & 4.51 & 6.08 & 6.38 & 13.6 & 6.59$^{+0.04}_{-0.05}$, 6.87$^{+0.07}_{-0.05}$ & 51.13$^{+0.07}_{-0.12}$, 50.67$^{+0.22}_{-0.27}$ & 4.86$\pm$0.07 \\
2015-08-11 & 2870 & 1.857$\pm$0.001 & 1.654$\pm$0.002 & 4.48 & 6.25 & 6.49 & 13.6 & 6.60$^{+0.02}_{-0.04}$, 6.90$^{+0.14}_{-0.18}$ & 51.09$^{+0.03}_{-0.16}$, 49.97$^{+0.68}_{-0.74}$ & 3.30$\pm$0.06 \\
2015-11-21 & 2921 & 1.872$\pm$0.001 & 1.714$\pm$0.002 & 6.95 & 9.0 & 9.3 & 19.2 & 6.57$^{+0.03}_{-0.05}$, 6.85$^{+0.05}_{-0.04}$ & 51.13$^{+0.07}_{-0.10}$, 50.76$^{+0.16}_{-0.21}$ & 5.17$\pm$0.06 \\
2016-02-06 & 2960 & 1.855$\pm$0.001 & \dots & 4.89 & 6.86 & 6.98 & 14.7 & 6.59$^{+0.04}_{-0.05}$, 6.84$^{+0.06}_{-0.04}$ & 51.17$^{+0.08}_{-0.12}$, 50.78$^{+0.18}_{-0.29}$ & 5.53$\pm$0.07 \\
2016-06-25 & 3030 & \dots & 1.669$\pm$0.001 & 8.84 & 11.2 & 11.4 & 23.6 & 6.57$^{+0.02}_{-0.02}$, 6.93$^{+0.02}_{-0.01}$ & 51.25$^{+0.03}_{-0.03}$, 50.97$^{+0.04}_{-0.05}$ & 7.38$\pm$0.06 \\
2016-07-07 & 3036 & \dots & 1.704$\pm$0.001 & 5.37 & 7.4 & 7.62 & 15.8 & 6.59$^{+0.03}_{-0.03}$, 6.91$^{+0.02}_{-0.03}$ & 51.19$^{+0.04}_{-0.06}$, 50.87$^{+0.10}_{-0.08}$ & 6.27$\pm$0.07 \\
2017-02-04 & 3142 & \dots & 1.695$\pm$0.001 & 8.33 & 11.0 & 11.0 & 21.5 & 6.45$^{+0.03}_{-0.04}$, 6.83$^{+0.01}_{-0.01}$ & 51.07$^{+0.05}_{-0.06}$, 51.09$^{+0.04}_{-0.05}$ & 6.75$\pm$0.06 \\
2017-05-20 & 3195 & 1.726$\pm$0.001 & 1.659$\pm$0.001 & 5.93 & 9.76 & 9.88 & 21.5 & 6.59$^{+0.02}_{-0.04}$, 6.87$^{+0.03}_{-0.05}$ & 51.19$^{+0.04}_{-0.09}$, 50.68$^{+0.20}_{-0.16}$ & 5.37$\pm$0.06 \\
2017-08-10 & 3236 & 1.729$\pm$0.001 & \dots & 4.4 & 6.48 & 6.54 & 13.6 & 6.54$^{+0.04}_{-0.06}$, 6.84$^{+0.07}_{-0.05}$ & 51.12$^{+0.09}_{-0.11}$, 50.73$^{+0.17}_{-0.30}$ & 4.86$\pm$0.07 \\
2017-11-20 & 3287 & 1.799$\pm$0.001 & 1.647$\pm$0.002 & 5.24 & 7.37 & 7.49 & 15.6 & 6.59$^{+0.03}_{-0.02}$, 6.92$^{+0.03}_{-0.02}$ & 51.24$^{+0.04}_{-0.04}$, 50.87$^{+0.06}_{-0.09}$ & 6.69$\pm$0.07 \\
2018-02-03 & 3325 & 1.755$\pm$0.001 & 1.646$\pm$0.001 & 13.2 & 13.1 & 13.3 & 27.4 & 6.59$^{+0.03}_{-0.01}$, 6.95$^{+0.03}_{-0.01}$ & 51.38$^{+0.04}_{-0.02}$, 51.11$^{+0.03}_{-0.09}$ & 10.10$\pm$0.07 \\
\hline
\end{tabular}
\end{small}
\end{center}
\renewcommand{\arraystretch}{1.}
\end{table*}

With the arrival of the large X-ray observatories, {\it XMM-Newton} and
{\it Chandra}, an interest in coronal cycles has been triggered.
The first discoveries of coronal cycles were made in several
stars with previously known chromospheric activity cycles: HD 81809
\citep{fav04,fav08,orl17}, 61 Cyg A \citep{hem06,rob12,bor16}, and
$\alpha$~Cen~B 
\citep{rob05,rob12,ayr09,dew10}, all of them in binary systems.
The two components of
$\alpha$~Cen and 61 Cyg are resolved spatially in X-rays, and
tentative cycles have been proposed, but not yet confirmed, for the
companions 61 Cyg B and $\alpha$~Cen~A \citep{rob12,ayr14} and for
Proxima~Cen \citep{war17}. All of them
have in common their   relatively old ages \citep[$>2$~Gyr,][]{barnes07}  and their long
rotation periods. Younger stars have shorter rotation periods and
higher levels of activity. The search for coronal cycles has shown
seasonal changes in EK~Dra and AB~Dor \citep{gud04,san07b}, but no
evidence of cycles after 35 years of X-ray monitoring of the very active
star AB~Dor \citep{lal13}. The shortest coronal cycle found to date is
that of 
\ihor\ \citep[1.6~yr, Sanz-Forcada et al. 2013, hereafter][]{san13},
being also the youngest and most active star for which a
coronal cycle has been identified. A very similar level of activity
is present in $\epsilon$~Eri, for which a $\sim$2.9~yr coronal cycle
has  recently been found (Coffaro et al., submitted).
\citet{war17} found a cycle in the dwarf M5.5 
star Proxima~Cen, with X-ray data that are partly consistent with the
photospheric counterpart, and an X-ray amplitude
$L_{\rm Xmax}/L_{\rm Xmin} \sim1.5-2$. These authors also find  a
positive correlation between the X-ray activity cycle amplitude and
Rossby number (assuming that Proxima~Cen and $\alpha$~Cen~A X-ray
cycles are real). 

%
\begin{figure}
  \centering
  \includegraphics[angle=270,width=0.45\textwidth]{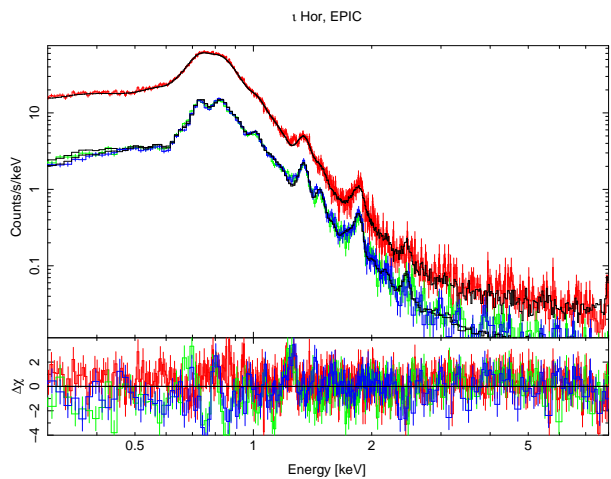}
  \caption{Spectral fit (black) for the summed {\it XMM-Newton}/EPIC-pn
    observation of \ihor\ for a total of 168, 228, and  234 ks in pn (red), 
    MOS1 (green), and  MOS2 (blue), respectively. The spectra include $1\sigma$ error bars.}\label{fig:epicspectotal} 
\end{figure}
%

\object{$\iota$ Hor} (HR 810, HD 17051) is an F8V/G0V star \citep{vauc08}
at a distance of 17.24$\pm0.16$~pc \citep{vanLeeuwen07}\footnote{GAIA
  DR2 reports a   distance of 17.33$\pm$0.02, but we used the HIPPARCOS
  measurement for easier comparison with earlier results.}. A giant
($M \sin i=2.48$~M$_{\rm J}$)
planet was found orbiting \ihor\ at 0.96 a.u. \citep{Kuerster2000,zec13}. In 
\citet{san13} we studied in
detail this moderately active star ($\log L_{\rm X}/L_{\rm bol}=-4.9$,
$\log R'_{\rm HK}=-4.6$), and established
for the first time the presence of a coronal
(X-ray) cycle of the same duration as the chromospheric 1.6~yr cycle
reported by \citet{met10}. With an age of $\sim$600~Myr
\citep[][and references therein]{san13}, it can be considered a
young solar analog. \ihor\ allows us to study the radiation environment at the
approximate age at which life appeared on Earth \citep[][and
  references therein]{cno07}.
It is the earliest
solar-like activity cycle discovered to date. A detailed study
of the corona of 
\ihor\ is of great importance not only for the knowledge of the
evolution of activity cycles with stellar age, but also for our
understanding of other coronal
parameters of the young Sun, such as coronal thermal structure or
abundances.
Given the few cases
of late-type stars with accurate measurements of their coronal and
photospheric abundances, we can put it in the context of the fractionation
effects that take place in the solar corona itself.
In this work we continue the analysis of the coronal cycle of
\ihor\ where we explore 
the existence of a second, longer  cycle, suggested by earlier
data in \citet{san13} and \citet{san16}.
This study complements the long-term study of the magnetic field of
\ihor\ whose initial results are presented in
\citet{alv18}. In that study we proposed that the time-evolution of
  chromospheric activity can be described by a superposition of two
periodic signals with similar amplitude, at
$P_1\simeq1.97$ and $P_2\simeq1.41$~yr. We also estimate a rotation
period of $P_{\rm rot}=7.70^{+0.18}_{-0.67}$~d, close to the proposed
values in the range 7.9--8.5~d from photometry and \ion{Ca}{ii}
variations \citep{vauc08,met10}.
A second longer  chromospheric cycle with irregular  amplitude is
suggested by the 
\ion{Ca}{ii}~H\&K data of \citet{flo17}, with a $P_{\rm
  cyc}\sim4.57$~yr, but the shorter 1.6~yr cycle was not identified in the
same data set.

\begin{table}
\caption{Coronal and photospheric abundances of \ihor\ (solar
  units\tablefootmark{a})}\label{tab:abundances} 
\tabcolsep 2.4pt
\begin{center}
\begin{small}
 \begin{tabular}{lrccccc}
\hline \hline
      & {FIP} &  \multicolumn{2}{c}{Solar photosphere} & \multicolumn{3}{c}{\ihor\ [X/H]} \\
  {X} & (eV)  &  Ref.$^a$ & (AG89)&  RGS\tablefootmark{b} & EPIC & Photosphere \\
\hline
Al & 5.98 & 6.45 & 6.47 &  0.29$\pm$0.19 & \dots         & 0.17$\pm$0.04 \\
Ni &  7.63 & 6.22 & (6.25) &  0.15$\pm$0.22 & \dots         & 0.14$\pm$0.05 \\
Mg &  7.64 & 7.60 & (7.58) & -0.09$\pm$0.28 & 0.13$\pm$0.03 & 0.17$\pm$0.05 \\
Si &  8.15 & 7.51 & (7.56) &  0.41$\pm$0.18 &-0.02$\pm$0.04 & 0.16$\pm$0.05 \\
Fe &  7.87 & 7.50 & (7.67) &      0.14      & 0.14$\pm$0.02 & 0.19$\pm$0.06 \\
S  & 10.36 & 7.12 & (7.21) & -0.09$\pm$0.35 & \dots         &  \dots \\
C  & 11.26 & 8.43 & (8.56) & -0.04$\pm$0.13 & \dots         & 0.21$\pm$0.08 \\
O  & 13.61 & 8.69 & (8.93) &  0.26$\pm$0.10 &-0.01$\pm$0.03 & 0.16$\pm$0.08 \\
N  & 14.53 & 7.83 & (8.05) &  0.24$\pm$0.13 & \dots         &  \dots \\
Ar & 15.76 & 6.40 & (6.56) &  0.78$\pm$0.39 & \dots         &  \dots \\
Ne & 21.56 & 7.93 & (8.09) &  0.13$\pm$0.15 & 0.00$\pm$0.06 &  \dots \\
\hline
\end{tabular}
\end{small}
\end{center}
\tablefoot{
\tablefoottext{a}{Solar photospheric abundances from \citet{asp09},
  adopted in this table, are expressed in logarithmic scale. 
Several values have been
updated in the literature since \citet[AG89]{anders}, also listed
for easier comparison.}
\tablefoottext{b}{Abundances measured using lines in RGS, except for
Si and Al that use only HST/STIS (colder) lines.}}
\end{table}

In Sect. 2 we describe the observations and  the analysis carried out, and  
we explain the results   in Sect. 3. We discuss the results in the
general framework of stellar coronae, and stellar activity in Sect. 4,
and present the conclusions in Sect. 5.

%
\begin{figure*}
   \centering
   \includegraphics[angle=90,width=0.9\textwidth]{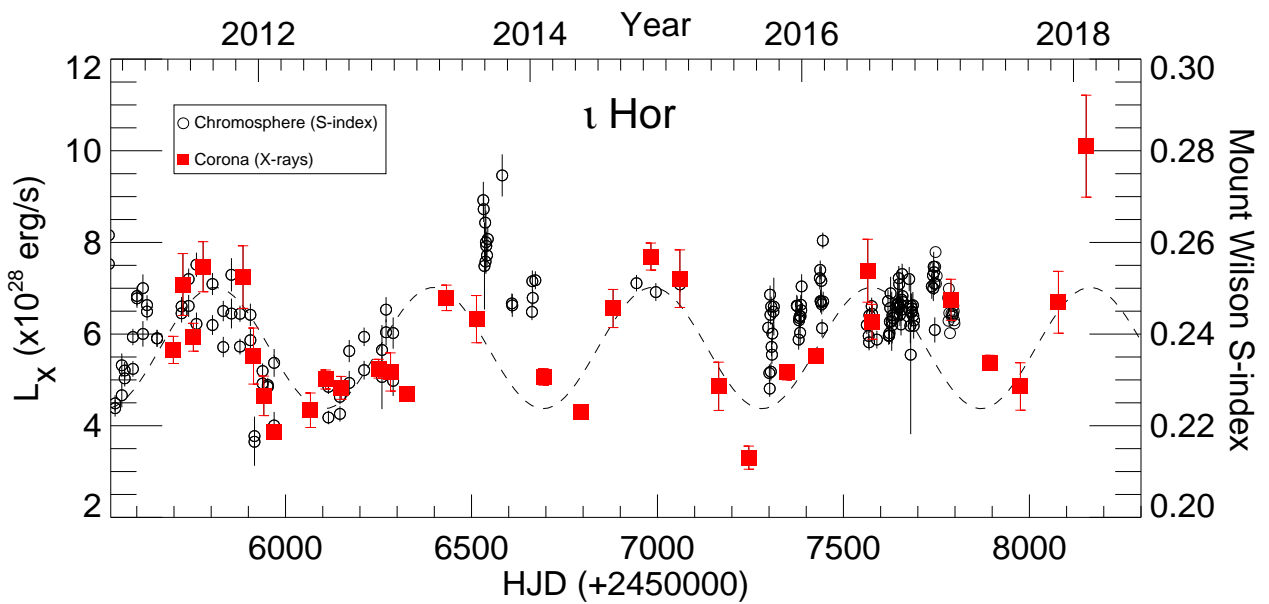}
   \caption{Time series of X-ray luminosity and
     chromospheric S-index for \ihor. The dashed line
     indicates the cycle calculated using just coronal X-ray data
     (period of 588.5~d).  The error bars of coronal X-ray luminosity are based
     on the standard deviation within each snapshot. Chromospheric data
     from \citet{san13} and \citet{alv18}.}\label{fig:cycle} 
\end{figure*}
%

\section{Observations}\label{sect:observations}
This work benefited from the information provided by different instruments covering X-rays, UV, and optical wavelength ranges, including photometric and spectroscopic information, as described below.

\subsection{XMM-Newton}\label{subsect:obs_xrays}
We  monitored the X-ray emission of \ihor\ between May 2012 and
February 2018 using {\it XMM-Newton} (proposal IDs 067361, 069355, 072247,
074402, 076383, 080338, P.I. J. Sanz-Forcada; and Director
Discretionary Time proposals 070198, 079018), for a total of 32
snapshots of 7--15~ks of total duration (Table~\ref{tab:xraylog} lists
the effective exposures from each instrument). {\it XMM-Newton} allows
the simultaneous use of all instruments on board.

\subsubsection{EPIC and RGS}\label{subsubsect:obs_xrays_instrum}
Data from the X-ray detectors were reduced following
standard procedures present in the Science Analysis Software (SAS)
package v16.1.0, removing the time intervals with high
background. The European Imaging Photon Camera (EPIC) PN and MOS
\citep[sensitivity range 0.1--15 keV and 0.2--10~keV, respectively,
  $E/\Delta E\sim20-50$,][]{tur01,str01} were used to monitor the long-
and short-term X-ray variability of the star. EPIC light curves
(Fig.~\ref{fig:lcxray}) and spectra were extracted for each
observation. All EPIC and  Reflection Grating Spectrometer (RGS) spectra were
fit using the ISIS package \citep{isis} and the Astrophysics Plasma
Emission Database \citep[APED,][]{aped} v3.0.9.
We summed all EPIC spectra, separately for PN, MOS1,  and MOS2, and we
simultaneously fit them to accurately calculate the
coronal abundance of Fe, O, Ne, Mg, and Si (Fig.~\ref{fig:epicspectotal}).
The summed EPIC spectra were fit
using a global 3-T model: $\log T_{1,2,3}$(K)=$6.15\pm0.02$,
$6.64\pm0.01$, $6.92\pm0.01$; $\log$~{\it
  EM}$_{1,2,3}$~(cm$^{-3}$)=$50.95\pm0.06$, $51.12\pm0.02$,
$50.81\pm0.02$, resulting in the abundances given in
Table~\ref{tab:abundances}. 
 A low value of interstellar medium (ISM)
absorption H column density of 3$\times10^{18}$~cm$^{-3}$ was adopted,
consistent with the fit to the overall spectrum, and the distance to
the source (17.2 pc)\footnote{The fit is not sensitive enough to
  provide an accurate value within the range $\sim
  1-7\times10^{18}$~cm$^{-3}$}. 
Global 2-T fits were used for the spectra of
each observation, fixing abundances to those of the summed spectrum.
Results of these fits are displayed in
Table~\ref{tab:xraylog}. The X-ray luminosity ($L_{\rm
  X}=7.93\pm0.01 \times10^{28}$ in the summed spectra) was calculated
in the usual {\em ROSAT} band 0.12--2.48 keV from the best-fit model
(Fig.~\ref{fig:cycle}).

\begin{figure*}
  \vspace{0.5cm}
  \centering
  \includegraphics[angle=270,width=0.9\textwidth]{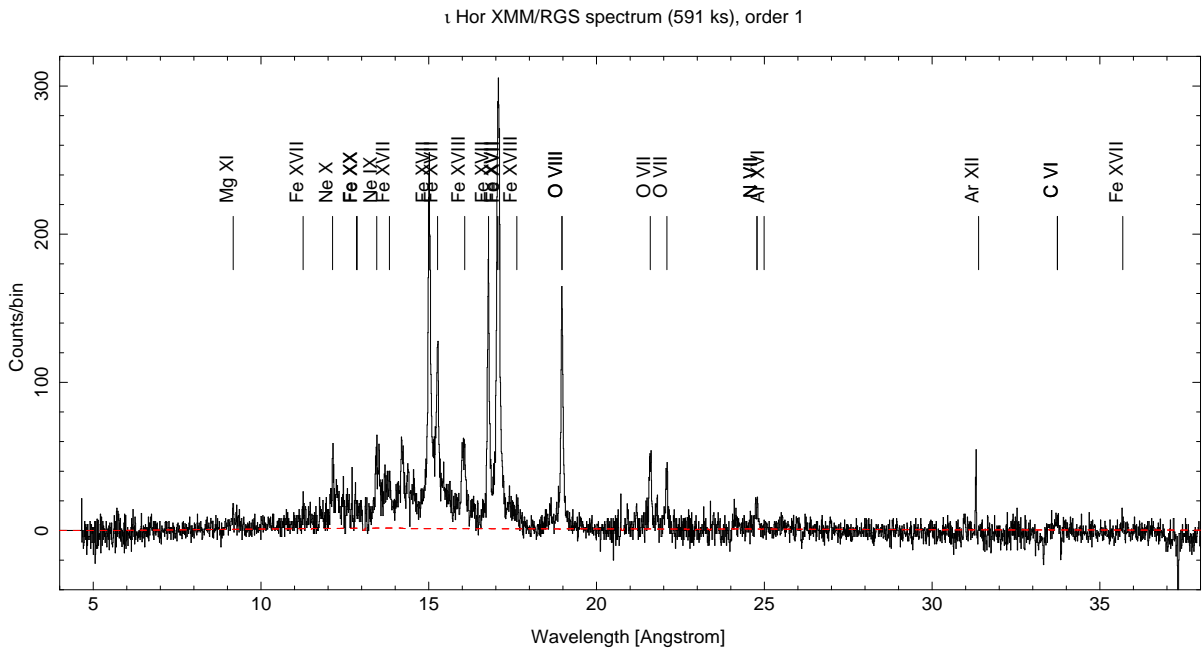}
  \caption{{\it XMM-Newton}/RGS (RGS1 + RGS2) summed spectrum of all observations, for a total of 591 ks. A dashed red line indicates the continuum below the emission lines.}\label{fig:rgs}
\end{figure*}

The RGS \citep[$\lambda\lambda \sim 6-38$~\AA,
$\lambda$/$\Delta\lambda\sim$100--500,][]{denher01} spectra were
summed to get an 
overall, time-averaged, high spectral resolution view of the
corona of \ihor\ (Fig.~\ref{fig:rgs}). In the case of RGS a
more complex procedure was employed to benefit from the rich
information provided by the measurements of mostly resolved spectral
lines. This information was used to construct an emission measure
distribution (EMD) multi-T model ($\Delta T=0.1$~dex) following
\citet{san03}. Spectral line fluxes with blend contributions, and
comparison with the line fluxes predicted by the model are listed
in Table~\ref{tab:rgsfluxes}. In this method line fluxes are measured
considering the local continuum predicted by an initial 2-T model (see
Fig.~\ref{fig:rgs} with local continuum at the end of the process),
convolving the response of the instrument with the lines. The measured
lines are compared with the fluxes predicted by an initial EMD
model. The observed ratios allow  the model to be changed in order to search for the
best ratios. The whole process is iterated to place a better continuum
until the results converge (Table~\ref{tab:emd},
Fig.~\ref{fig:emd}). The [Fe/H] abundance is fixed to the value
determined with EPIC,
where the continuum is more sensitive to this parameter than the RGS
continuum. This method also allows   the abundances of some
elements in the corona to be measured, as shown in Table~\ref{tab:abundances}
and Fig.~\ref{fig:abund}.

%

\subsubsection{The Optical Monitor}\label{subsubsect:obs_om}
The {\it XMM-Newton} Optical Monitor (OM) was used to observe \ihor\ either
in ``Image'' or ``Time Series'' modes in the UV,
with the UVW2 ($\lambda=2120 \AA$, $\Delta\lambda=500 \AA$) and UVM2
($\lambda=2310 \AA$, $\Delta\lambda=480 \AA$) filters in the different
campaigns.
We used the OM data as processed in the official {\it XMM-Newton} pipeline
products. The UV flux densities in these bands
during the different exposures are listed in Table~\ref{tab:xraylog}.
By averaging all OM observations for a given filter we found a mean 
observed flux density 
of $F_{\rm UVM2} = (1.84 \pm 0.06) \ 10^{-12}\, \rm erg\,s^{-1} cm^{-2} \AA^{-1}$
for the UVM2 band and
$F_{\rm UVW2} =(1.67 \pm 0.01) \ 10^{-12}\, \rm erg\,s^{-1} cm^{-2} \AA^{-1}$ for the UVW2 band.  
In Sect.~\ref{subsect:discussion_uv} we use these values together with
the flux densities predicted from photospheric model atmospheres to
calculate the chromospheric contribution to the UV flux of $\iota$\,Hor.

\subsection{HST}
We acquired {\em Hubble Space Telescope} ({\em HST}) observations on
2018 Sep 3 through {\it HST} Proposal ID 15299 (P.I. J. D. Alvarado-G\'omez),
using the Space Telescope Imaging Spectrograph (STIS) with the E140M
grating (sensitivity range $1150-1740~\AA$, $\lambda/\Delta
\lambda=11500-17400$). 
Two observations of 3141~s of exposure time each were summed using
IRAF STIS package software, to get a better quality spectrum.   
We then measured the line flux of lines formed in the $\log T$(K)$=4.1-5.6$
range to extend 
our EMD analysis towards the transition region and upper chromosphere
of \ihor. Two lines, \ion{C}{ii} 1334.535~\AA\ and \ion{Si}{iv}
  1402.7704~\AA, were affected by the ISM
  absorption in less than $\sim20\%$. In these two cases the 
line flux was measured by fitting a Gaussian to the emission line. All
line wavelengths and measured fluxes are listed in
Table~\ref{tab:stisfluxes}.

\subsection{TESS}\label{subsubsect:obs_tess}
We explored the data from NASA's {\em Transiting Exoplanet Survey Satellite}
(\textit{TESS}) mission \citep{ric15} 
to calculate the rotational period of $\iota$\,Hor and to assess its
photometric activity. 
The short-cadence (2 minutes) data of \ihor\ were collected by
\textit{TESS} in sector~2 and sector 3, starting on 2018 Aug 23
and 2018 Sep 20, respectively, and were made publicly available with the 
first data release in December 2018 and January 2019. The target pixel
file, which 
consists of 19737 and 19692 cadences for sectors 2 and 3,
  respectively,  was downloaded from the    Barbara 
A. Mikulski Archive for Space Telescopes (MAST) Portal. The light curves were
obtained using the \textit{Kepler}GO/lightkurve code \citep[version:
  1.0b13, August 2018;][]{lightkurve18}.  
To create the apertures that were used to extract the light curves from
the target pixel files, all cadences from one sector were summed
into a single 
image. Then all the pixels of this combined image that have a value
within a certain percentage of the maximum flux value were used in the
extraction aperture. Ten different apertures with 0.1, 0.2, 0.3, 0.4,
0.5, 1, 2, 3, 4, and  5\% of the maximum flux value were tested.  The
light curve with the lowest standard deviation was chosen as the final
one. The aperture that was used to extract the final light curve
consisted of all pixels with flux values higher than 0.2\% and 0.4\%
  of the maximum value for sectors 2 and 3 light curves, respectively.
 
\textit{TESS} assigns a quality flag to all measurements. We removed
all the flagged data points except  ``Impulsive outlier'' and ``Cosmic ray
in collateral data'' (bits 10 and 11) while extracting the light
curve. The final light curve, which is shown in Fig.~\ref{TESS_LC},
consists of 31770 data points. It shows a quasi-periodic variation
with an amplitude below the percent level. Such variability is typical
for rotating star spots.

\section{Results}\label{sect:results}

The variety of observations analysed give a rich number of results. We
present first the results from the X-rays photometric analysis, then the
combination of X-rays and UV spectroscopy, UV photometry, and optical
light curves.

\subsection{X-ray variability}\label{subsect:results_xrayvariab}
 Short-term  X-ray variability $(f_{\rm max}-f_{\rm min})/f_{\rm
  min}\la 40\%$ is observed during almost all  
epochs, but no large flares were identified.
Small flares lasting up to $\sim$1~hr could be present in August
2011 and February 2018 (Fig.~\ref{fig:lcxray}).
The X-ray luminosity $L_{\rm X}$ displays a remarkably stable
variability pattern. The X-ray
amplitude ($L_{\rm X max}$/$L_{\rm X min}$) is 2.3 in the standard
$0.12-2.48$~keV band, except for the 
last maximum of the series (February 2018), with a value three times larger
than the minimum reached in August 2018 (Fig.~\ref{fig:cycle}). This
maximum could be affected by a larger scale flaring activity;
we note the increasing flux level and the two flares in the
light curve of this interval (Fig.~\ref{fig:lcxray}). 

We 
used the X-ray luminosity values from all observations to calculate a
periodogram using the  generalized Lomb-Scargle (GLS) software,
implemented by \citet{zec09}. 
The Lomb-Scargle (LS) periodogram is shown in
Fig.~\ref{fig:xray_periodogram}. The only significant peak is at
$587$~d. 
The X-ray light curve phase-folded with this period is shown in
Fig.~\ref{fig:phasefolded}, overlaid with the sinusoid corresponding
to this highest peak in the periodogram. The mean X-ray luminosity and the
half-amplitude of the X-ray cycle obtained from the sinusoid are
$\langle L_{\rm x} \rangle = 5.8 \times 10^{28}\,{\rm erg s^{-1}}$
and $\Delta L_{\rm x} = 1.4 \times 10^{28}\,{\rm erg s^{-1}}$,
respectively, i.e., $\Delta L_{\rm x} / \langle L_{\rm x} \rangle =
24$\,\%. 
The error associated with the period was found through Monte Carlo
simulations: each value of the X-ray luminosity in the light curve was
considered as the mean of a normal distribution, and from this a set
of 1000 normally distributed random numbers was generated for each
data point. We thus obtained 1000 simulated X-ray light curves on
which we performed the LS analysis. The
distribution of the resulting periods is a Gaussian.
We adopt the peak of this distribution and its standard deviation,
 $588.5\pm 5.5$\,d, as the period of the X-ray cycle and its uncertainty. 
The {\em TESS} and {\em HST}
  observations took place near the expected minimum of the activity cycle.

\begin{table}
\caption{Emission measure distribution of \ihor.}\label{tab:emd}
\renewcommand{\arraystretch}{1.3}  
\begin{center}
\begin{small}
\begin{tabular}{cc|cc}
\hline \hline
       {log~$T$ (K)} & {EM (cm$^{-3}$)\tablefootmark{a}} & {log~$T$ (K)} &
       {EM (cm$^{-3}$)} \\
\hline
4.0 & 51.15: & 5.8 & 48.35$^{+0.20}_{-0.30}$ \\
4.1 & 51.05: & 5.9 & 48.45$^{+0.30}_{-0.30}$ \\
4.2 & 50.85: & 6.0 & 48.75$^{+0.30}_{-0.20}$ \\
4.3 & 50.65$^{+0.30}_{-0.10}$ & 6.1 & 48.95$^{+0.20}_{-0.30}$ \\
4.4 & 50.45$^{+0.10}_{-0.30}$ & 6.2 & 49.50$^{+0.30}_{-0.20}$ \\
4.5 & 50.25$^{+0.10}_{-0.30}$ & 6.3 & 49.85$^{+0.20}_{-0.10}$ \\
4.6 & 49.95$^{+0.10}_{-0.30}$ & 6.4 & 49.55$^{+0.30}_{-0.20}$ \\
4.7 & 49.85$^{+0.20}_{-0.20}$ & 6.5 & 49.65$^{+0.30}_{-0.20}$ \\
4.8 & 49.75$^{+0.40}_{-0.10}$ & 6.6 & 49.75$^{+0.40}_{-0.20}$ \\
4.9 & 49.70$^{+0.30}_{-0.10}$ & 6.7 & 50.40$^{+0.20}_{-0.10}$ \\
5.0 & 49.75$^{+0.10}_{-0.20}$ & 6.8 & 50.70$^{+0.10}_{-0.10}$ \\
5.1 & 49.75$^{+0.10}_{-0.20}$ & 6.9 & 49.95$^{+0.40}_{-0.10}$ \\
5.2 & 49.45$^{+0.10}_{-0.30}$ & 7.0 & 50.05$^{+0.10}_{-0.10}$ \\
5.3 & 49.25$^{+0.10}_{-0.30}$ & 7.1 & 49.35$^{+0.20}_{-0.20}$ \\
5.4 & 48.90$^{+0.10}_{-0.30}$ & 7.2 & 48.70$^{+0.20}_{-0.30}$ \\
5.5 & 48.45$^{+0.20}_{-0.20}$ & 7.3 & 47.75: \\
5.6 & 48.25$^{+0.20}_{-0.30}$ & 7.4 & 47.55: \\
5.7 & 48.25$^{+0.20}_{-0.20}$ & 7.5 & 47.35: \\
\hline
\end{tabular}
\end{small}
\end{center}
\tablefoot{
\tablefoottext{a}{Emission measure (EM=log $\int N_{\rm e} N_{\rm H}
  {\rm d}V$), where $N_{\rm e}$
  and $N_{\rm H}$ are electron and hydrogen densities, in
cm$^{-3}$. Errors provided are not independent
between the different temperatures, as explained in \citet{san03}}}.
\renewcommand{\arraystretch}{1.}
\end{table}

\subsection{Emission measure distribution and abundances}
\begin{figure}
  \centering
  \vspace{-0.5cm}
  \includegraphics[width=0.45\textwidth]{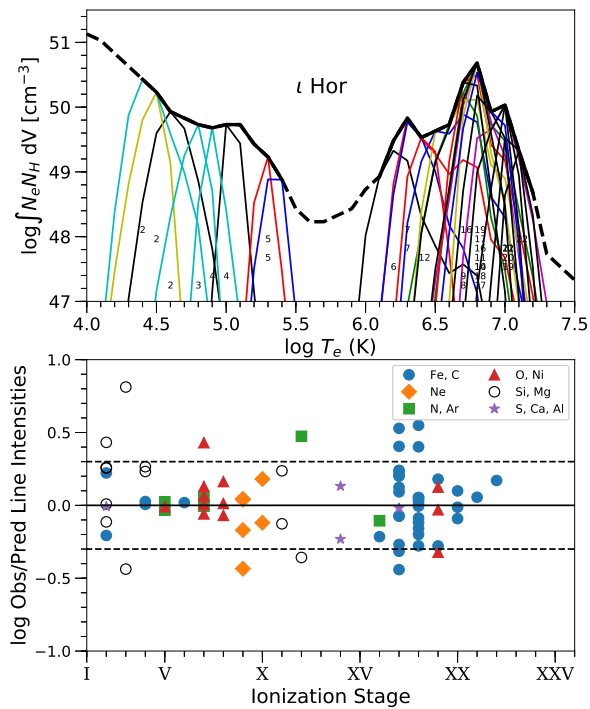}
  \caption{{\em Upper panel}: Emission measure distribution   of
    \ihor\ calculated using the summed RGS spectrum. The thin lines
    represent the relative contribution function for each ion (the
    emissivity function multiplied by the EMD at each point). The small
    numbers indicate the ionization stages of the species. {\em Lower
      panel}:  Shown are the observed-to-predicted line flux ratios for
    the ion stages in the upper figure. The dotted lines denote a
    factor of 2.}
  \label{fig:emd}
\end{figure}

The EMD (Fig.~\ref{fig:emd},
Table~\ref{tab:emd}), determined as explained above from the summed
RGS and STIS\footnote{Although STIS data were taken near the
  expected minimum of the cycle, in September 2018, the amplitude of the
  cycle is small. We can assume that the difference in emission level
  with the average RGS spectrum will have little impact on the
  determination of EMD and abundances.}
spectra of all observations, shows a
moderately active star with a main peak at $\log T$~(K)=6.8 and a
lower peak  at $\log T$~(K)=6.3 resembling that of low-activity stars
like the Sun or 
$\alpha$~Cen~B \citep{orl00,san11}. Some material at temperatures
above 10 MK is 
also found, but less abundant than in very active stars such
as AB Dor. The  corona of \ihor\ has similar characteristics to that
of $\epsilon$~Eri \citep{san04} or $\kappa$~Cet \citep{cno07}.
Loop models show that the  highest EM is found at the maximum
temperature of the loop \citep[e.g.,][]{gri98,car06}. Therefore, the EMD
of \ihor\ can be interpreted as the result of a combination of loops with their
maximum temperatures at the two mentioned peaks. The lower temperature
is not identified 
in the EPIC spectra because the global fitting methods yield the
dominant temperatures, and material at $\log T$~(K)=6.6 (found in EPIC
fits) has an EM similar to the average of the peak at $\log
T$~(K)=6.2--6.4.
Thanks to the UV lines observed with STIS, our coronal model
  (the EMD) is extended to the transition region and upper
  chromosphere. This allows us to estimate
the flux in different extreme
ultraviolet (EUV) bands useful for evaluating the impact on
exoplanet atmospheres, following the method explained in
\citet{san11}: $L_{100-920 \AA}=1.7\times10^{29}$\,erg\,s$^{-1}$, $L_{100-504
  \AA}=5.5\times10^{28}$\,erg\,s$^{-1}$.

%

The EPIC spectra of each of the snapshot observations were fit in
order to calculate the X-ray luminosity.
The results of the fits were used to identify the origin of the
cycle. A visual inspection of Fig.~\ref{fig:emtcycle} suggests
that the emission measure of the hot component is the most sensitive
of the four variables ($\log T_{1,2}$ and $\log EM_{1,2}$) to the
activity cycle (especially at the minima of 2014, 2015, and 2017).
We  correlated these variables with $L_{\rm X}$,
showing that the correlation is more direct with the emission
measure (correlation factor $r=0.82, 0.68$, and probability
  value\footnote{$p$ defines the probability that the data 
  are not correlated
$p=10^{-8}, 1.7\times 10^{-5}$}, respectively, with the
cool and hot components) than with the temperatures ($r=0.26, 0.56$,
$p=0.15, 0.00095$). The cycle is thus  modulated by the amount of
material in the corona. 

\begin{figure}
  \centering
  \includegraphics[width=0.45\textwidth]{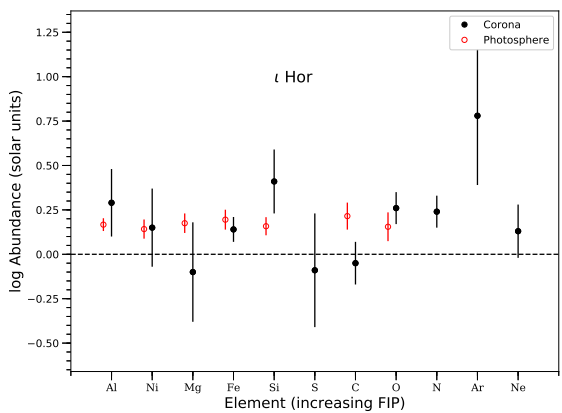}
  \caption{Coronal abundances (from {\it XMM-Newton}/RGS) and photospheric
abundances    \citep{gon07}  of \ihor\ in solar units, using
    \citet{asp09} as  reference.} 
  \label{fig:abund}
\end{figure}

Coronal abundances were also measured while fitting the
EMD\footnote{A reliable coronal Ca abundance could not
  be determined. The Ca abundance is based on only one line, which
  could also be affected by blending by with a nearby \ion{Cr}{xv} line.} using
the atomic models of APED, which are based on solar photospheric
abundances by \citet{anders}.
We then updated them to the reference values of
\citet{asp09}. Both are shown in Table~\ref{tab:abundances} for
completeness, together with the  \ihor\ photospheric abundances. Several
authors report photospheric abundances of
\ihor\ \citep{bon06,bia12,sot18}. The values for most elements are
quite similar in the literature, although \citet{bon06} calculates a
lower [Fe/H]$\sim$0.08. We
compare our results with those of \citet{gon07}, which cover the
longest list of elements common to our list of coronal elements,
including oxygen and carbon, those with largest first ionization
potential (FIP)  in the list (Fig.~\ref{fig:abund}).
Neither coronal nor photospheric abundances of \ihor\
deviate substantially from solar photospheric values. We thus
conclude that there are no effects related to FIP in the corona of
this star.

\subsection{UV variability}\label{subsect:results_uvvariab}
The UV measurements obtained with the {\em XMM-Newton} OM define a time series of
the same length as the X-ray data (i.e., 7 years). The long-term UV
light curve of $\iota$\,Hor is displayed in Fig.~\ref{fig:cycle_uv}.
Some of the observations included measurements with both the UVM2 and
the UVW2 filters. We
used the flux ratio of the star in the two bands ($f_{\rm
  UVM2}/f_{\rm UVW2} \sim0.9$) from these observations to scale
down the UVM2 measurements 
for a better display of data from both UV filters in Fig.~\ref{fig:cycle_uv}.
The variation in the flux density is limited to $\sim$14\% in the
UVM2 filter and to only $\sim$4\% in the UVW2 band.
No periodic pattern is visible in the UV light curve. We
calculated a periodogram with the GLS software, using the combined data of
both filters. No significant periodicity is found
(Fig.~\ref{fig:uv_periodogram}). 

%
%

\subsection{Starspot modulation}\label{subsect:results_rotation}
\begin{figure}
  \vspace{0.5cm}
  \centering
  \includegraphics[width=6cm,angle=270]{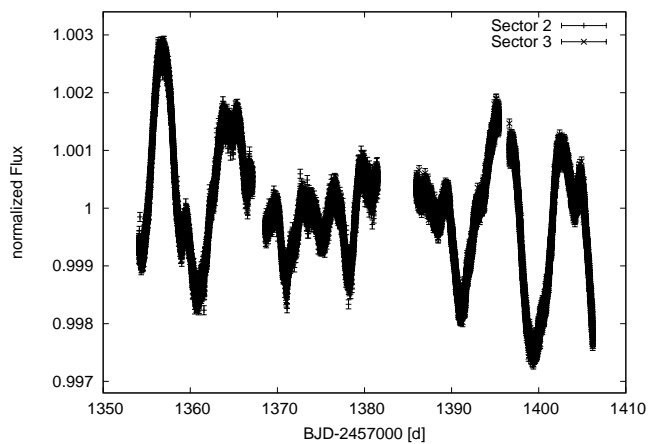}
  \caption{Two-minute cadence light curve of $\iota$\,Hor observed by
    \textit{TESS} in 2018 September.}
  \label{TESS_LC}
\end{figure}

\begin{figure}
  \centering
  \includegraphics[width=0.48\textwidth]{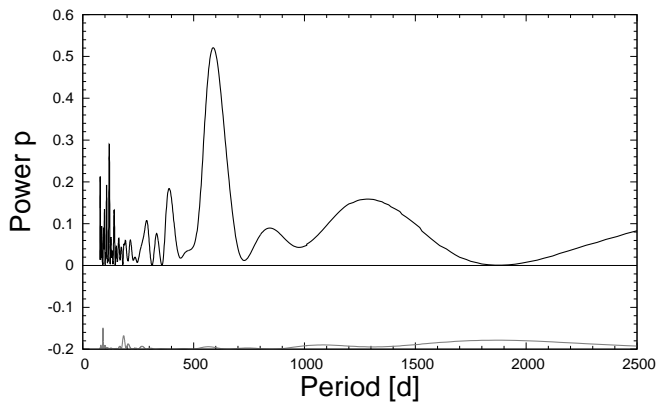}
  \caption{Generalized Lomb--Scargle power spectrum (top
    panel) and window function (bottom panel) of the long-term X-ray
    light curve of \ihor. The significant peak is at $587$\,d.}
  \label{fig:xray_periodogram}
\end{figure}

\begin{figure}
  \centering
  \includegraphics[width=0.45\textwidth]{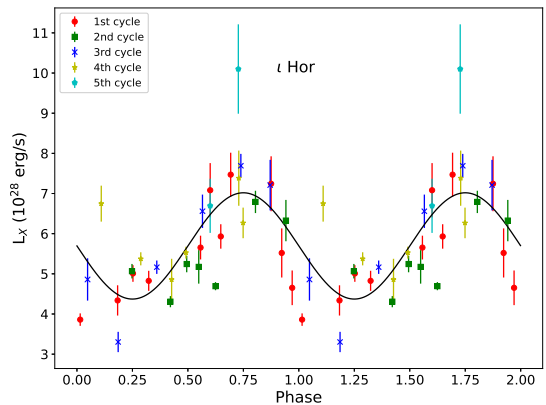}
  \caption{Phase-folded long-term X-ray light curve of \ihor\,
    based on the period of $588.5$~d, obtained from the GLS periodogram 
    analysis  described in Sect.~\ref{subsect:results_xrayvariab}.
    Superposed is the corresponding sine curve.} 
  \label{fig:phasefolded}
\end{figure}

\begin{figure*}
   \centering
   \includegraphics[width=0.45\textwidth]{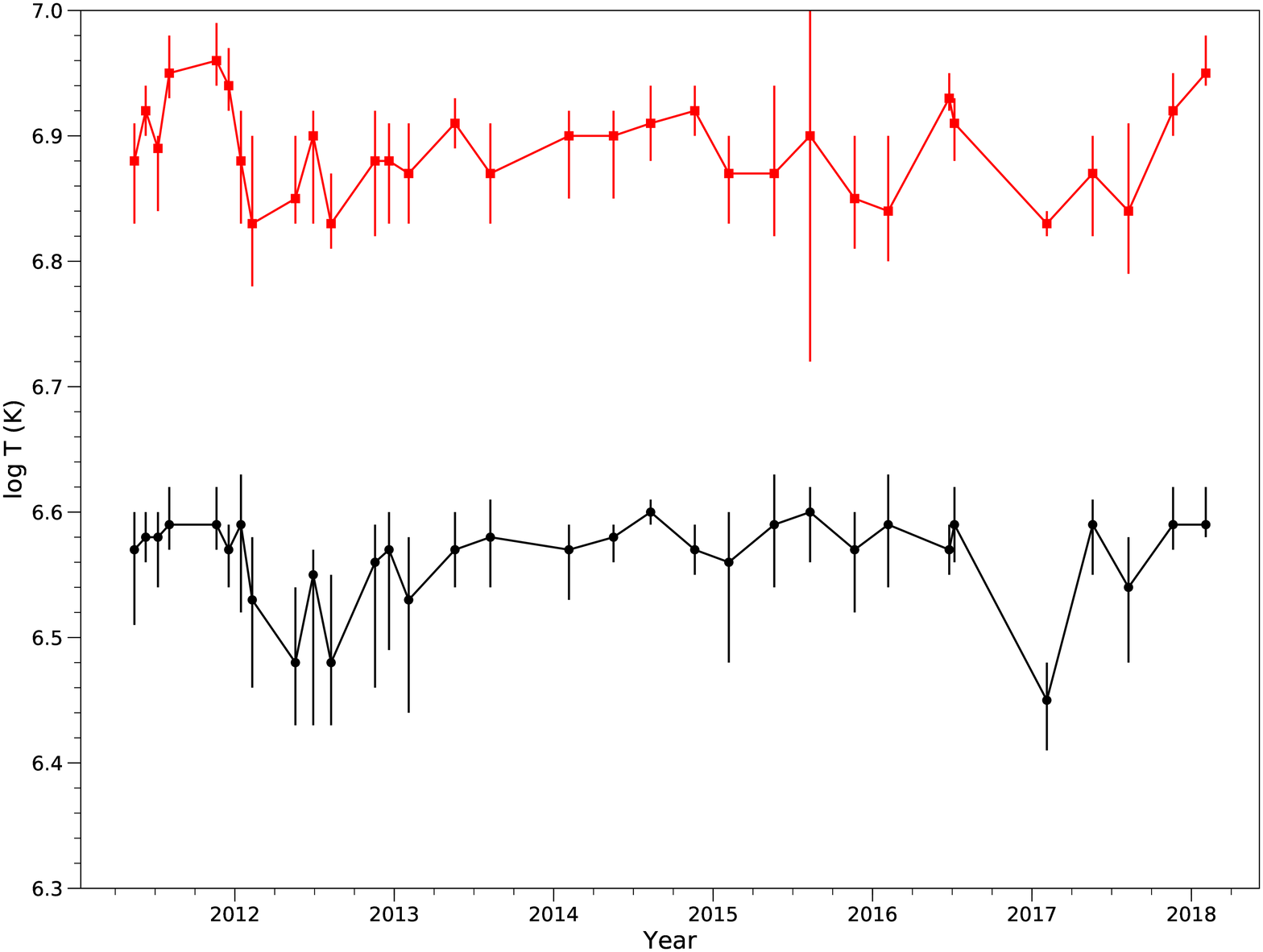}
   \includegraphics[width=0.45\textwidth]{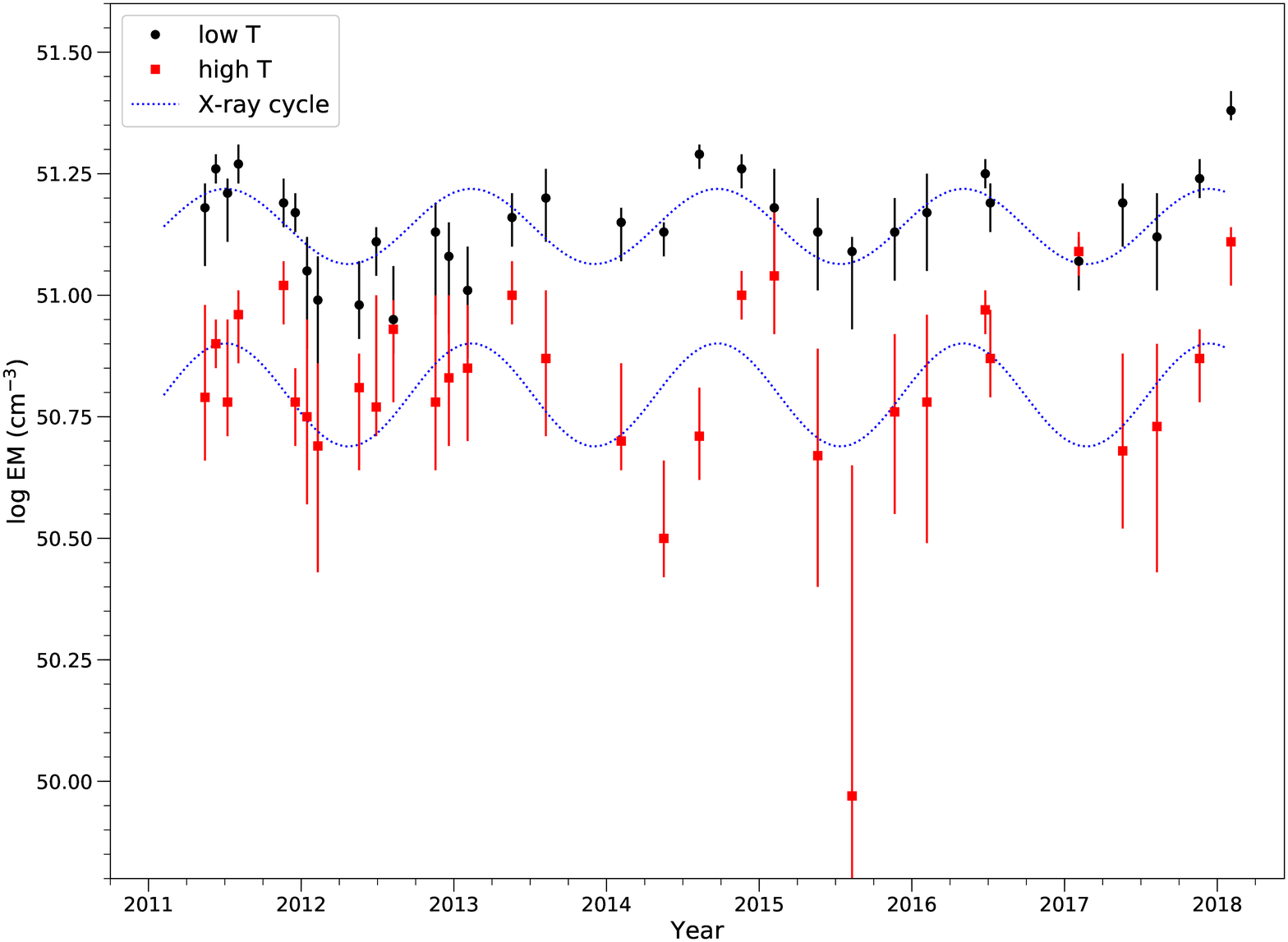}
   \caption{Time series of temperature (left panel) and emission measure
     (right panel) of the 2-T fits to the EPIC spectra of
     \ihor. The pattern of the X-ray light curve fit is shown for
     comparison.}\label{fig:emtcycle}  
    \end{figure*}

\begin{figure}[t]
   \centering
   \vspace{-0.5cm}
   \includegraphics[width=0.45\textwidth]{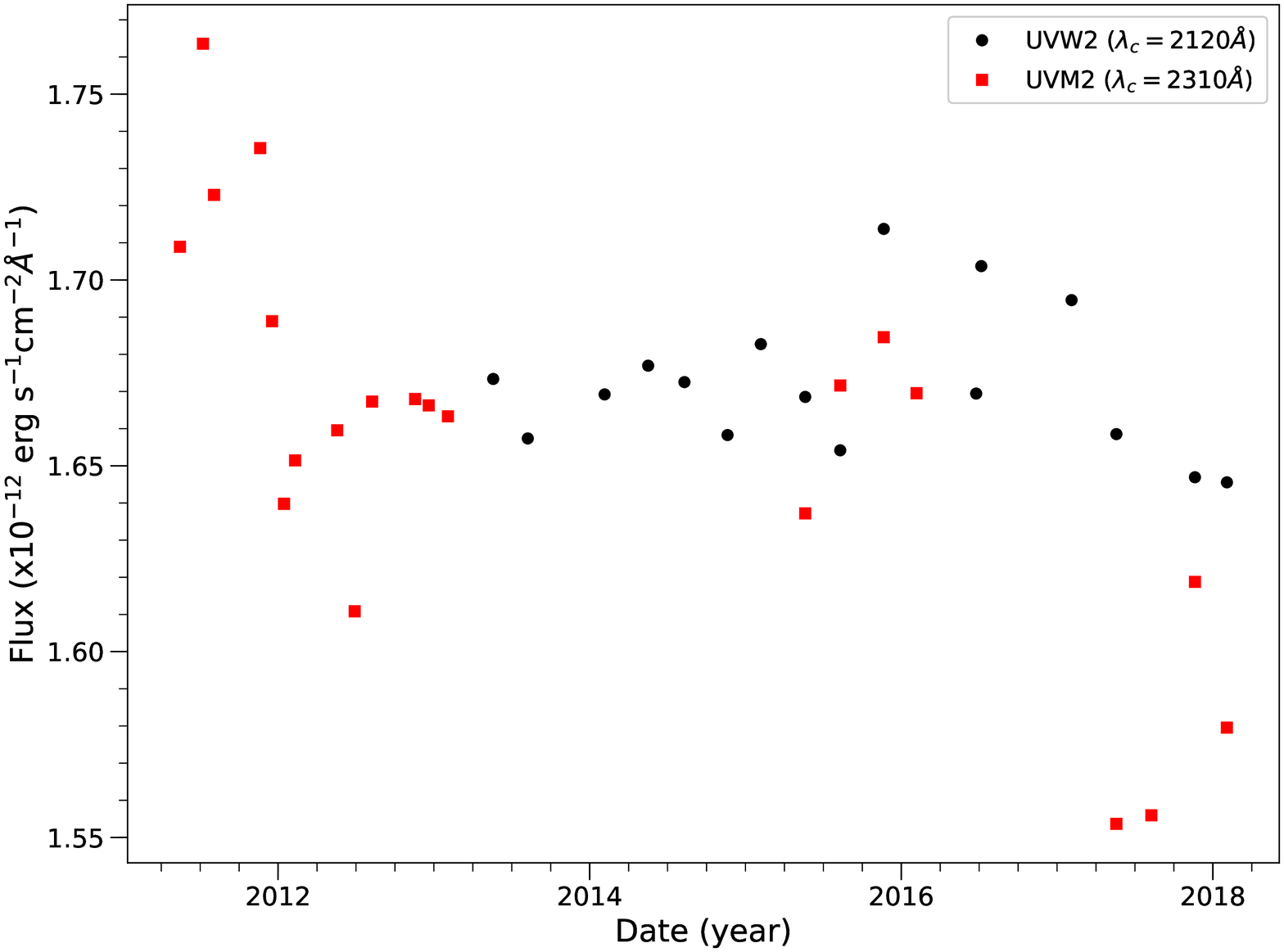}
   \caption{{\it XMM-Newton}/OM time series in the two UV filters. Data from
     UVM2 filter are multiplied $\times 0.9$ for a better display. No
     activity cycle is evident in the UV. }\label{fig:cycle_uv} 
    \end{figure}

We searched for the rotation period of $\iota$\,Hor using the same GLS routine on the {\it
  TESS} light curve  that we had applied to the
X-ray time series in Sect.~\ref{subsect:results_xrayvariab}
First, we used GLS on each {\it TESS} sector individually.
As this GLS implementation can only deal with up to 10000
data points we binned the data 
of each sector by a factor of 3. The highest peaks in the
periodograms correspond to a rotation period of $8.43\pm0.02$~d and
$7.95\pm0.02$~d for sectors 2 and 3, respectively
(Fig.~\ref{period_TESS}). Then we repeated 
the GLS analysis for the full {\it TESS} light curve (both sectors
combined). The data had to be binned by a factor of 4. The
periodogram clearly shows two significant peaks at
$P_{\mathrm{1}}=7.718\pm0.007$~d and
$P_{\mathrm{2}}=9.47\pm0.02$~d (Fig. \ref{period_TESS}). Our GLS
analysis and the 
double-humped structure of the light curve suggest two
dominating spots. As can be seen from Fig.~\ref{TESS_LC} the
morphology of the light curve changes from one rotational cycle to
the next indicating spot evolution. The observed change in period
between observations of Sectors 2 and 3 might be due to differential
surface rotation combined with changes in the spot latitude. A
quantitative assessment of this effect is beyond the scope of this
work. We adopt
the mean of the values ($P_{\mathrm{1}}$, $P_{\mathrm{2}}$)
from each of the two sectors periodograms, $P_{\mathrm{rot}}=8.19\pm0.26$~d. 

To our knowledge, we have derived here the first photometric measurement of the 
rotation period for $\iota$\,Hor. In Table~\ref{tab:rotation} we
compare our result to measurements presented previously in the
literature based on different diagnostics. As can be seen, all these
values cluster around $\approx 7.0-8.5$\,d. However, the LS
periodogram obtained from the {\it TESS} light curves provides a much higher
significance than the other methods \citep[see, e.g., Fig.~6 in][]{alv18}. 

We combined the {\it TESS} measurement of $P_{\rm rot}$ with the
published values for the projected rotational velocity \citep[$v
  \sin{i} = 6.0\pm0.5$\,km/s;][]{alv18} and the stellar radius
\citep[$R=1.17\pm0.04\,{\rm R_\odot}$,][]{fuhrmann17}, and we determined the
inclination of \ihor\ to be $i = 56^\circ$ (range $49^\circ-65^\circ$
taking into account  the uncertainties).

\subsection{White-light flares}\label{subsect:results_flares}
We also searched for flares in the {\it TESS} light curve of $\iota$\,Hor. 
Our flare analysis procedure is based on the routine used in
\citet{ste16}. With an iterative process of boxcar smoothing of the
light curve and removing $3\sigma$ outliers we created a final
smoothed light curve, that was interpolated to all data points of the
original input light curve. All points that lie $3\sigma$ above the
final smoothed light curve were flagged. All groups of at least three
consecutive flagged points were assigned as potential flares. 

To validate potential flares as flare candidates, five criteria have to
be fulfilled: (i) the flare does not happen right before or after a
data gap, (ii) the maximum flux value is significant with at least 
$3\sigma$, (iii) the flux ratio between the flare maximum and the
last flare point is $\geq$3, (iv) the difference between the second-to-last and the last flare point is smaller than the standard deviation
of the cleaned light curve, and (v) a flare template suggested by
\citet{dav14} fits the data better than a linear fit through all flare
points. 

   We could not detect any flares in the {\it TESS} light curve of
$\iota$\,Hor (i.e., there were no groups of at least three consecutive outlying
points). Therefore, the flare validation procedure described in the
previous paragraph could not be applied. We  present it here
because in Sect.~\ref{superflares} we analyze  flares in a
comparison sample to evaluate the possibility of \ihor\ hosting
(rare) superflares.

%

\section{Discussion}\label{sect:discussion}

The number of results obtained yield 
  different topics to be discussed: the X-ray activity cycle, 
  \ihor\ and the early evolution of the solar corona, the
  observed pattern in the coronal abundances, the net UV emission
  coming from the chromosphere, and the analysis of \ihor\ in the
  context of superflare stars.

\subsection{X-ray activity cycle}
The chromospheric behavior in \ion{Ca}{ii} H\&K was used to 
identify the activity cycle \citep{met10}, but later observations
indicate a more erratic behavior of the chromospheric cycle in the
period between 2013 and 2017, also resulting  in a small amplitude of 
the S-index, as shown in
Fig.~\ref{fig:cycle}. \citet{alv18} explain the observed behavior as
the superposition of two periods of 1.97 and 1.41~yr of similar
amplitudes. The origin of these two periods remains unclear.

The observed coronal cycle gives rise to an intriguing question: Why is
the chromospheric cycle of \ihor\  substantially more irregular than
its coronal 
counterpart? The combination of the chromospheric and coronal data
initially pointed to a modulation of the 1.6~yr cycle with a longer
term periodicity \citep{san13,san16}.
However, the inspection of the
chromospheric and the coronal longer data set now available
indicates that coronal
cycle is quite regular in duration and amplitude, while the chromospheric
cycle is irregular and seems to have faded since late 2013. A geometrical
effect was suggested by \citet{san13} to explain the irregularities
observed earlier in the chromospheric cycle, just before the X-ray
monitoring started.
With a stellar rotation axis inclination of $i\sim 56^\circ$ we
might be biased  
towards viewing the activity in one of the hemispheres, but with
irregularities introduced by the different visibility of the
more occulted
hemisphere. This effect should be less evident in X-rays, since X-ray
emission originates in coronal material, more extended over the stellar
photosphere, in addition to being optically thin. The observed
behavior in \ihor\ seems to support this idea.

%
%

\subsection{\ihor\ as a young solar analog}
The RGS spectroscopic analysis reveals a very similar coronal structure to
that of $\kappa$~Cet (G5V, $\log L_{\rm X}/L_{\rm bol}=-4.4$,
Sanz-Forcada in prep.). Both have a similar age  
\citep[$\sim 600$~Myr,][and references therein]{lac99,san13}
and can be considered a sort of proxy of the young Sun,
given their solar-like
spectral types. The age of both stars corresponds to the estimated
solar age when the 
earliest known forms of life appeared on Earth \citep{cno07}. The
coronal properties of \ihor\ are of special interest to calibrate the
effects of high-energy photons and particles on the early Earth.
The stellar wind properties are expected to vary during the activity cycles
\citep{oran13,alv16}. Effects of the solar cycle on the Earth's
atmosphere include the variation of cosmic rays arriving to the
atmosphere, the change in the size of the
exosphere, or the thermosphere cooling fluctuations
\citep[e.g.,][]{fri91,sve16,mly18}. The influence of activity cycles
could introduce a modulation in the atmosphere of the planet thorough
these effects, although on a different scale from that detected on the Earth
given the different length and amplitude of the cycle in \ihor\ and
the Sun.
It has been proposed that activity cycles can also act as weather seasons
for biological processes in exoplanets under some circumstances
\citep{mul18}.

%
%

%
%

\subsection{Coronal abundances}
\citet{per15} used the EPIC-pn spectrum to measure the coronal
abundances of a few elements in \ihor, which were then compared with the
photospheric values in the literature. Our superb RGS
combined spectrum allows for a more accurate measurement of the
coronal abundances (Table~\ref{tab:abundances},
Fig.~\ref{fig:abund}). The comparison we make with the photosphere of
\ihor\ shows no trend related to the FIP.
Our results differ from those of \citet{per15} who find an FIP effect
(elements with low FIP are enhanced in the corona with respect to the
photosphere) in \ihor\ based on the [Fe/O] ratio. However, they calculate a
very low oxygen coronal abundance, [O/H]=$-0.45\pm0.05$, quite
different from our value from EPIC ([O/H]=$-0.01\pm 0.03$) or RGS
([O/H]=$0.26\pm0.10$) spectra. This discrepancy in the EPIC results with
\citet{per15} could be due to our better
statistics, which allow us to use a 3-T model to fit the spectrum,
thus with a better
sampling of the region where oxygen lines are formed.
Our result also contradicts the conclusion made by \citet{woo10},
followed by \citet{lam15}, in
which FIP-related effects in dwarfs are correlated with spectral
type, using only stars with $\log L_{\rm X}$(erg s$^{-1}$)$<29$.
This would imply that \ihor\ (G0V) would have  an FIP
effect, while M dwarfs would suffer an inverse FIP
effect. The authors use solar photospheric values for their comparison
in the dwarf M stars,  instead of stellar photospheric abundances. 
In the case of GK dwarfs they use in the comparison the Mg and Si
  as low FIP elements rather than Fe,
which is the low-FIP element
with the best determined abundances.
Finally, the exclusion of high-activity stars in the
sample of the above-mentioned FIP studies sheds some doubts on
whether the FIP (or inverse FIP) effect is 
related to spectral type or to stellar activity.

A question that remains open is whether the coronal abundances remain
constant during the  activity cycle. A recent work reveals that solar
coronal abundances may vary along the activity cycle
\citep{bro17}. The quality of our data does not allow us to search for
variations in the abundances between the different observations, an
exercise that would be reliable only with high resolution
(RGS-like) spectra with enough statistics.

\subsection{UV chromospheric emission}\label{subsect:discussion_uv}
For late-type stars the observed UV flux is a combination of photospheric and
chromospheric contributions. To examine what fraction of the UV
emission measured by the OM is from the chromosphere of
\ihor,\ we calculated the chromospheric excess flux density
($F_{\rm UV,exc}$) as the difference between the UV observed flux
density 
and the photospheric flux density ($F_{\rm sy}$) predicted by
atmosphere models for the respective UV waveband (i.e.,
$F_{\rm UV,exc}= F_{\rm UV} - F_{\rm sy}$). 
We carried out this analysis for the two filters of the {\em XMM-Newton}
OM used during our campaigns and for published {\em GALEX} measurements
\citep{shk13}. The {\em GALEX} near ultraviolet filter (NUV)
observation was discarded 
because \ihor\ was saturated 
according to the detector limits reported by \citet{mor07}.

In order to determine the expected photospheric flux, we first
extracted all the published photometric data of $\iota$ Hor from the
catalogs in VizieR. Using  the online tool {\it Virtual
  Observatory SED Analyzer} \citep[VOSA;][]{bayo08}
we then fit these data with the model BT-Settl-CIFIST \citep{allard14,
  baraffe15}. As input parameters the SED fitting procedure
requires the effective
temperature ($T_{\rm eff}$), the logarithm of the surface gravity
($\log g$), and the metallicity. We fixed the gravity and the
metallicity values to $\log g = 4.5$ and [Fe/H]$=0$. These are
the values in the spectral grid that are the closest to the literature
values for \ihor\ according to \citet{fuhrmann17}. In the best fit we
obtained $T_{\rm eff} = 6000 \pm 50$\,K, consistent with the effective
temperature given by \citet{fuhrmann17}.
We then calculated the photospheric flux density as 
$F_{\rm sy} = \int F_{\rm mod}(\lambda) T_{\rm filter}(\lambda) d\lambda$.
Here, $F_{\rm mod}(\lambda)$ is the flux density of the synthetic
BT-Settl-CIFIST spectrum\footnote{The synthetic model spectra are all
  available at the France Allard web page:
  http://perso.ens-lyon.fr/france.allard/},
with $T_{\rm eff}$ equal to the best-fit parameter $6000$\,K, and  
$T_{\rm filter}(\lambda)$ is the normalized transmission curve of the 
respective UV filter. 
Since the synthetic spectrum is available in terms of surface flux
density, we used $R_\ast = 1.17$ and stellar distance to get the flux
density at Earth.

In Table~\ref{tab:UV_flux_and_excess} we provide the observed and the
excess flux densities for the three UV filters with reliable data.
The errors stated in
Table~\ref{tab:UV_flux_and_excess} were propagated from the
dilution factor $\left(R_\ast/D\right)^2$.
The results show an expected increase in the chromospheric excess for
decreasing wavelengths. Finally, 
the chromospheric far-ultraviolet (FUV) excess of \ihor, $L_{\rm
  FUV,exc}$= $(1.140 \pm 0.005) \times 10^{29}$ erg s$^{-1}$, is
consistent with the relation between X-rays and FUV chromospheric emission
reported in \citet{piz19}.

\subsection{Photometric activity}\label{subsect:discussion_tess}
One result of the analysis of the photometric data collected by NASA's
\textit{Kepler} mission was the detection of flare activity on
solar-like stars, which before had rarely been seen. A number of these
stars show flares with energies more than ten times larger than the largest
known solar event ($10^{32}$ erg). The \textit{Kepler} data revealed
hundreds of sun-like stars with these so-called ``superflares''
\citep[e.g.,][]{mae12,not13,shi13}. 
In a theoretical work \citet{air16} concluded that superflares, and the
coronal mass ejections associated with them, could serve as a
potential catalyst for the origin of life on the early Earth.
\citet{kar16} explored the relation between superflare stars and their
chromospheric emission for a sample of stars with
$T_{\rm eff}=5100-6000$~K. In that work the distribution of
stars with superflares 
peaks at a \ion{Ca}{ii} H\&K S-index $\sim 0.26$, well above the stars
with no flares ($\sim 0.19$). According to this distribution \ihor\ is a
good candidate to be a superflare star.

%

As described in Sect.~\ref{subsect:results_flares}, in our analysis 
of the \textit{TESS} data of \ihor\ we could not find any flare
events, but since the data span in total only an observing time of
$\sim$52\,d, 
we wanted to test the hypothesis of \ihor\ being a superflare star,
comparing it with a sample of {\it Kepler} superflare stars.
If $\iota$\,Hor had a  flare rate similar to that of the stars in our
\textit{Kepler} sample ($0.03-0.23\,{\rm N_{\rm flares}/day}$; see
Appendix~\ref{superflares}) we would expect to detect between $1.4$
and $11.1$ (average: $3.8$) flares in the $\sim 48$\,d of actual
\textit{TESS} observations. 
To quantify the probability of no flaring events during the {\it TESS}
observations of $\iota$\,Hor, we simulated the incidence of flares in a
putative observing time span of $t_{\rm obs} = 1$\,Myr. Using the
average flare rate $0.08\,{\rm N_{Flares}/day}$ of our Kepler sample
(see Sect.~\ref{superflares}) and assuming this flare rate to be
constant for the whole $t_{\rm obs}$, we obtained $29.2 \times 10^6$
flares, which  we randomly injected at times between $t=0$ and $t_{\rm
  obs}$. Then we measured the time lags, $\delta t$, between
consecutive flares. Finally, we calculated the fraction of all $\delta
t$ that are larger than the duration of $\iota$\,Hor's {\it TESS}
light curve. 
We found that the probability of no flares being observed in the time
spanned ($52.1$\,d)
is $1.55$\%. If we also consider the gap of $4.4$\,d
between sectors~2 and~3 where possible flares could go
unnoticed, the probability for zero flares in the observed light curve
increases to $2.20$\%. There is, therefore, a small probability that
$\iota$\,Hor is a superflare star, but no such event has been discovered yet.
We also note  that our estimate ignores that the sensitivity for flare
detection is probably somewhat lower for {\em TESS} with respect to {\em
  Kepler} as a result of its redder bandpass ($\lambda\lambda
6000-10000 \AA$ vs $\lambda\lambda 4200-8800 \AA$).

\section{Conclusions}\label{sect:conclusions}
The X-ray observations of \ihor\ confirm a robust coronal cycle of
$P=588.5\pm5.5$~d, 
with an amplitude of 2.3 in the 0.12-2.48~keV X-ray band.
The coronal cycle is modulated by the amount of loops present in
  the corona.
The effective temperature and age of this star make it a good 
representation of the Sun at an age of $\sim 600$\,Myr, when life
appeared on Earth. 
The understanding of the high-energy properties of such young
solar analogs is thus of paramount importance for assessing the
variable irradiation to which a proto-Earth is exposed.
Chromospheric cycles show a direct relation between rotation and cycle
periods. The X-ray (coronal) cycle comparison between the ``early
Sun'' (\ihor) and the present-day Sun reveal that cycles in the more
  evolved state are
longer in duration than cycles of the younger Sun, but they also have  a
larger amplitude 
in X-rays. It is expected that the influence of the X-ray cycle in the
atmosphere of planets is also greater now. Given the short duration of
early cycles, we can speculate that their effects mimic 
 seasonal weather patterns for biological processes, as proposed by
\citet{mul18}.

The X-ray observations of \ihor\ cover four complete and
consecutive coronal cycles, more than observed on the Sun. This has
allowed us to study cycle-to-cycle variability, and we noticed 
an irregular behavior of the 
chromospheric signal, but a more regular coronal modulation. We
propose that this disagreement is due to a 
geometrical effect. Given the inclination of $\sim 56^\circ$ (see
  below) we do not
observe in a symmetric view the northern and southern hemispheres of the
star. While this may have some effects on the chromospheric emission
depending on the spot coverage of each hemisphere, which may be out
of phase, the coronal signal comes from more extended material and
should be less sensitive to viewing effects.

The spectroscopic analysis of the corona of \ihor\ shows a medium
activity level star with an abundance pattern similar to the solar
photosphere, with no FIP-related effects. This pattern
contradicts the earlier results measured in the literature using
low-resolution X-ray spectra. It also breaks the dependence of the 
FIP effect with spectral type proposed by \citet{woo10}.

%
\begin{table}
\begin{center}
\caption{Rotation period estimates for \ihor.}
\label{tab:rotation}
\begin{tabular}{ccc}
\hline \hline
$P_{\rm rot}$ (d) & diagnostic & reference\tablefootmark{a} \\ 
\hline
$8.19 \pm 0.26$ & photometric time series & this work \\
$7.7$  & S-index & (1) \\
$7.88$ & longitudinal field & (1) \\
$7.03$ & radial velocity & (1) \\
$7.9$ & chromospheric activity & (2) \\
$7.9-8.5$ & S-index & (3) \\
$7.9-8.4$ & radial velocity & (4)\\
\hline
\end{tabular}
\end{center}
\tablefoot{
  \tablefoottext{a}{(1) - \citet{alv18}, (2) - \citet{Saar97}, (3) -
    \citet{met10}, (4) - \citet{boi11}}} 
\end{table}

%
\begin{table}
\caption{Observed UV flux densities and chromospheric excess\tablefootmark{a}}\label{tab:UV_flux_and_excess}
\tabcolsep 3.pt
\begin{center}
\begin{small}
\begin{tabular}{lccc}
  \hline \hline
Filter & $\lambda_{\rm eff}$ & $f_{\rm \lambda, obs}$ & $f_{\rm \lambda,exc}$ \\ 
& ($\AA$) & (erg\,s$^{-1}$\,cm$^{-2} \AA^{-1}$) & (\%)\\
\hline
{\em GALEX}/FUV & $1516$ & $(1.29 \pm 0.03) \times 10^{-14}$ & $ 71 \pm 3 $ \\
OM/UVW2 & $2120$ & $(1.67 \pm 0.01) \times 10^{-12}$ & $ 39 \pm 4  $ \\
OM/UVM2 & $2310$ & $(1.84 \pm 0.06) \times 10^{-12}$ & $ 20 \pm 7  $ \\
\hline
\end{tabular}
\end{small}
\end{center}
\tablefoot{\tablefoottext{a}{Chromospheric excess values derived as described in the main text for a photosphere model with $T_{\rm eff} = 6000$\,K.}}
\end{table}

We have presented the first photometric measurement of the rotation
period for \ihor\ using {\it TESS} data. Our value is consistent with
earlier measurements from spectroscopic data (e.g., S-index, radial
velocity, longitudinal magnetic field). Combining our value for
$P_{\rm rot}$ ($8.19 \pm 0.26$\,d) with historical measurements of
$v\sin{i}$ and stellar radius, we derived a new estimate for the
inclination, $i = 56^\circ$. No white-light flares are detected in the
{\it TESS} light curve, but our evaluation of the flare amplitudes of
{\em Kepler} superflare stars has shown that the possibility of
occasional superflares on $\iota$\,Hor cannot be
excluded. Superflares would likely have strong effects on the early
atmosphere of the Earth, especial if they yielded coronal mass
  ejections \citep[but see][and references  therein]{alv18b}.

The UV emission does not follow the cyclic behavior
detected in the X-ray light curve, and its variability amplitude
is limited to  $\sim14$\% at most.
We used the UV spectral energy distribution to  
quantify for the first time the chromospheric broadband UV flux of 
$\iota$\,Hor, after subtracting the photospheric contribution
to the SED.

The multi-wavelength study of long-term variability in other stars of
similar or younger age will tell us how general  the case of
\ihor\ is, and whether it is the age of the first activity cycles in stars
like the Sun. The study of coronal cycles in other stars at different
ages is also of interest to understand how the amplitude of the
  X-rays emission in the cycle evolves with stellar age.

\begin{acknowledgements}
We acknowledge the anonymous referee for the useful comments that
helped improve the manuscript. This work has made use of the {\it
  XMM-Newton} and {\it TESS} 
space telescopes and archives, operated by ESA and NASA, respectively, and
{\it HST}, managed by both agencies.
We acknowledge Norbert Schartel for the observations granted as
{\it XMM-Newton} Director Discretionary Time (DDT).
We acknowledge Nancy Brickhouse and Adam Foster for their help in
the interpretation of the problem with the Ca abundance.
JSF acknowledges support from the 
Spanish MINECO through grants AYA2008-02038,
  AYA2011-30147-C03-03, and AYA2016-79425-C3-2-P.
MC acknowledges financial support from the \textsl{Bundesministerium
  f\"{u}r Wirtschaft und Energie}  
through the \textsl{Deutsches Zentrum f\"{u}r Luft- und Raumfahrt
  e.V. (DLR)} under  grant number FKZ 50 OR 1708.
JDAG was supported by grants from Chandra (GO5-16021X) and HST
(GO-15299).
This research has made use of the SVO Filter Profile Service
(http://svo2.cab.inta-csic.es/theory/fps/) supported by the Spanish
MINECO through grant AYA2017-84089
\end{acknowledgements}


\begin{appendix}
\section{Supplementary material}
%
\begin{figure}
  \centering
  \includegraphics[clip,width=0.49\textwidth]{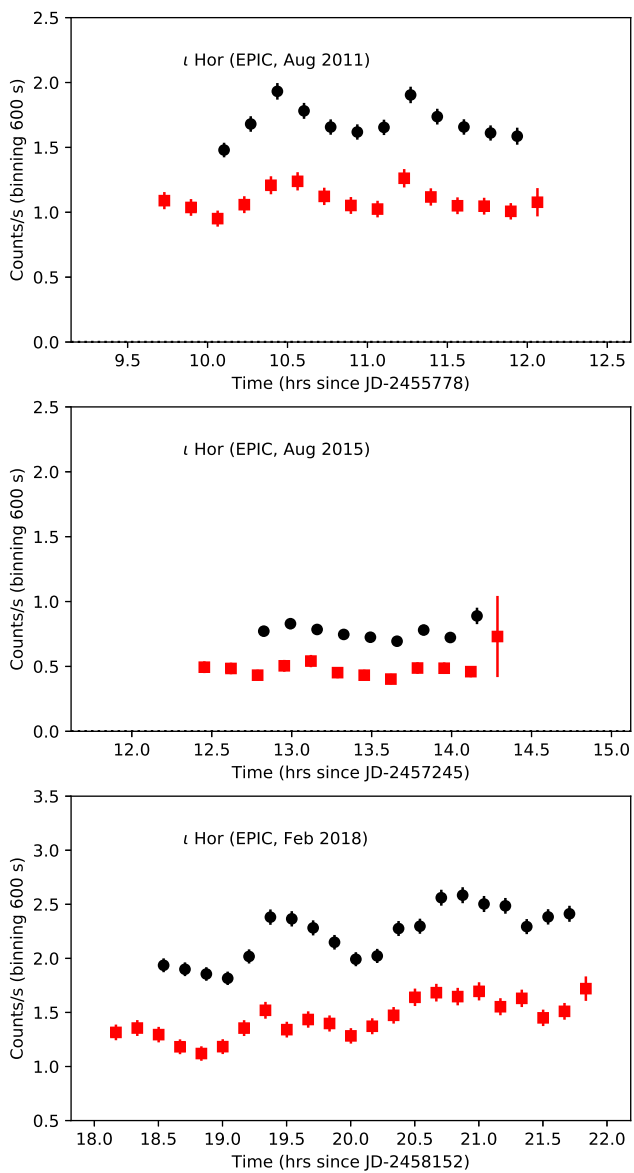}
  \caption{Time series for the XMM-Newton EPIC (black, pn + MOS1 + MOS2) and MOS (red, MOS1+MOS2) observations of
    \ihor\ during the lowest activity level and highest flaring activity,
    with 1$\sigma$ error bars. MOS data are scaled up ($\times 2$)
    for a better display with EPIC data. MOS observations  start
    earlier than EPIC-pn exposures. The count rate scale of
      the February 2018 observation is larger than the others.}
  \label{fig:lcxray}
\end{figure}
%

%
\begin{figure}
  \centering
  \includegraphics[width=0.48\textwidth]{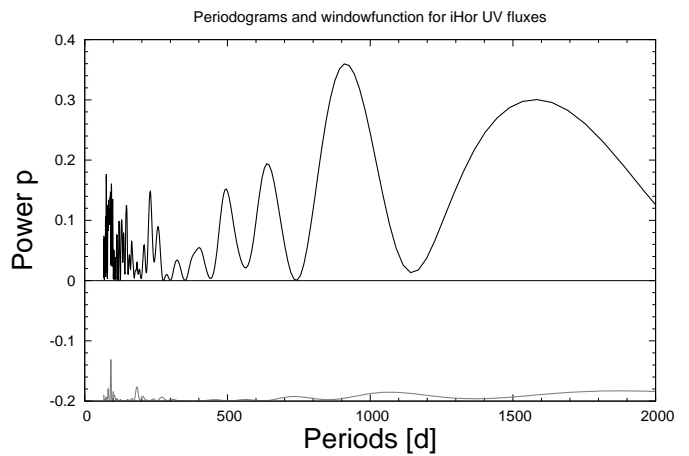}
  \caption{Generalized Lomb--Scargle power spectrum (top
    panel) and window function (bottom panel) of the long-term OM/UV
    light curve of \ihor. The highest peak, with low significance, is
    at 909\,d.} 
  \label{fig:uv_periodogram}
\end{figure}
%

%
\begin{figure*}
\centering
  \includegraphics[angle=270,width=0.33\textwidth]{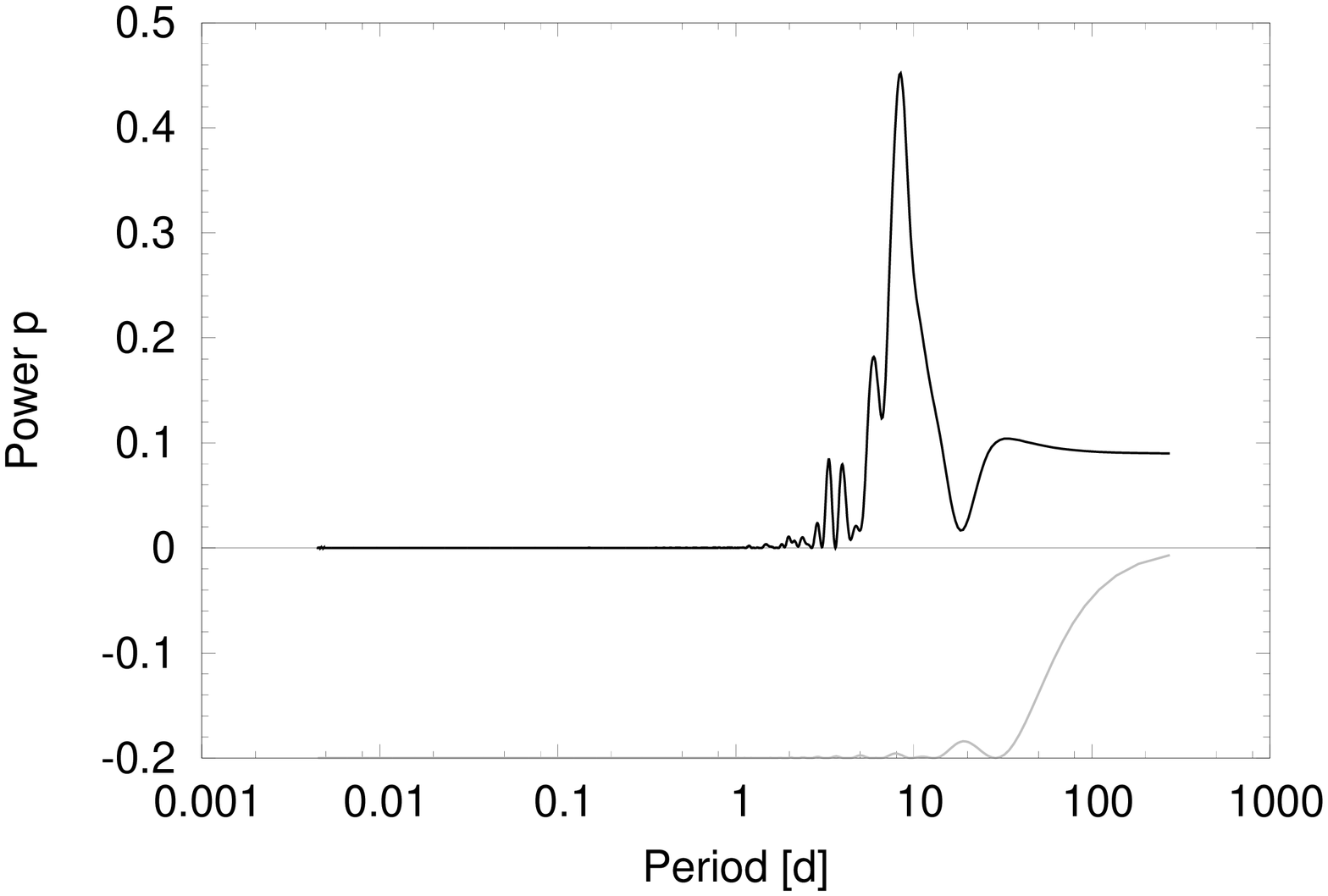}
  \includegraphics[angle=270,width=0.33\textwidth]{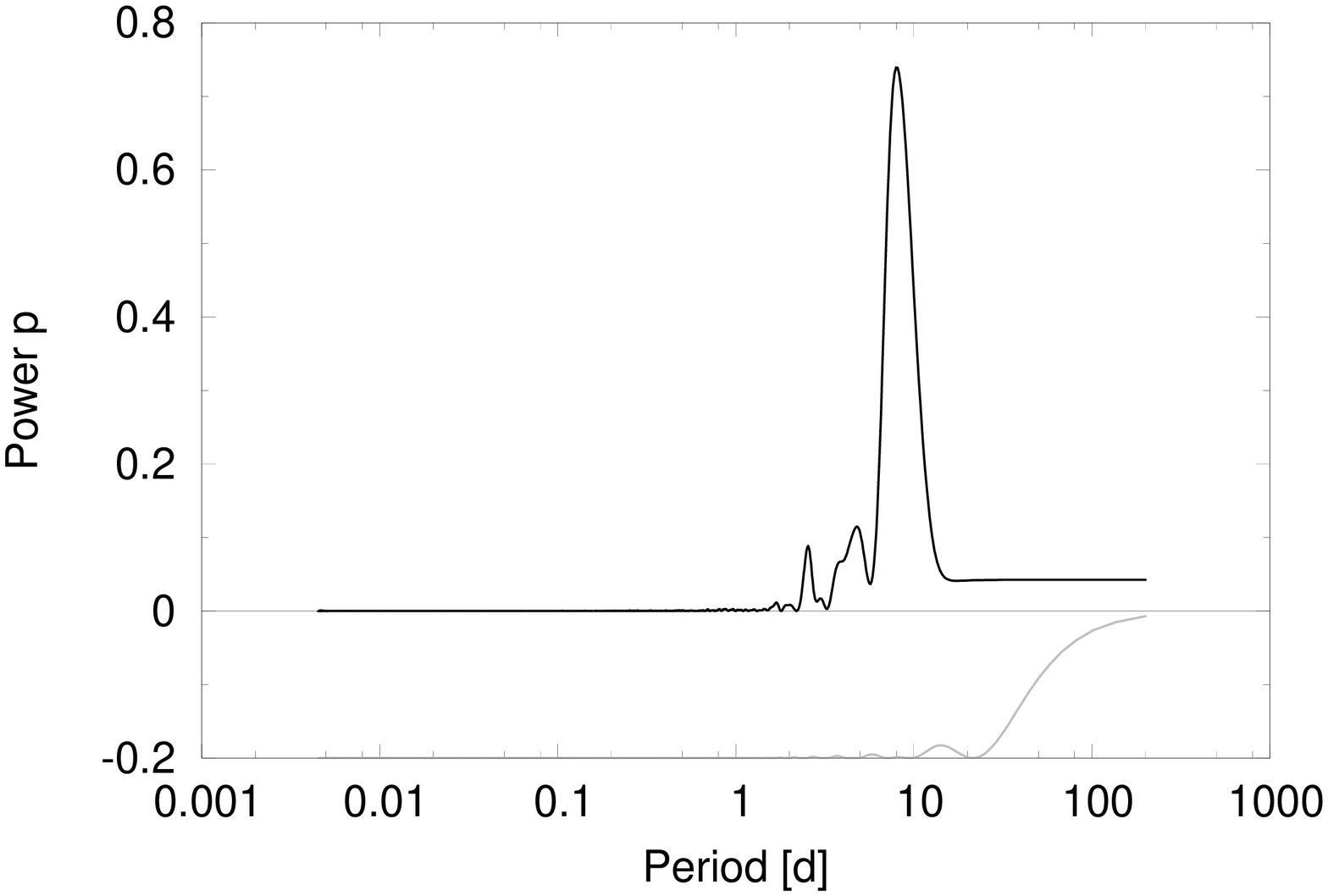}
  \includegraphics[angle=270,width=0.33\textwidth]{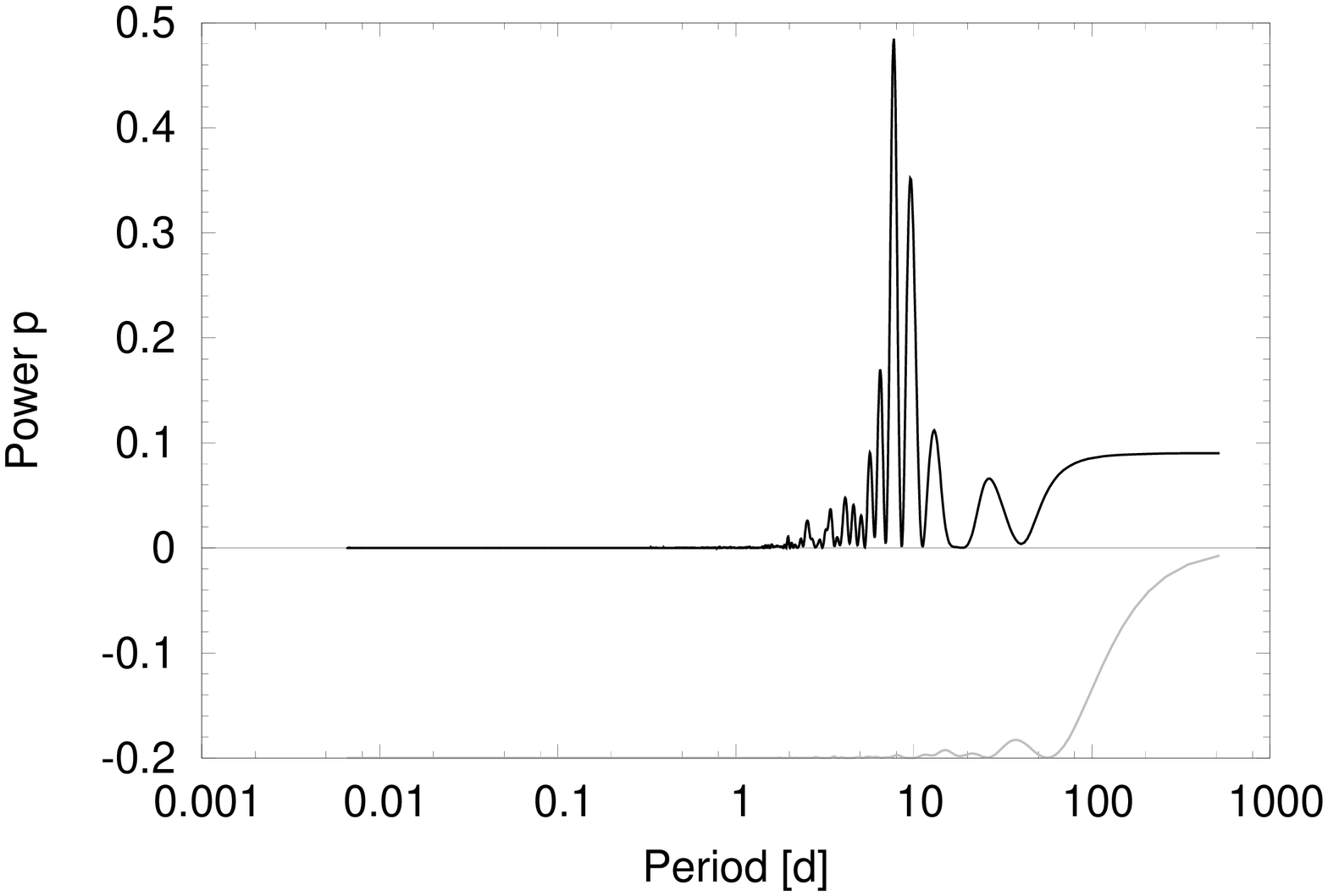}
  \caption{Generalized Lomb--Scargle power spectrum (top panels) and
    window function (bottom panels) for the \textit{TESS} light curve
    of \ihor\ in  Sector 2 ({\it left}), Sector 3 ({\it center}), and
    combined data of sectors 2 and 3 ({\it right}). The highest peaks are
    at a period of $P_{\rm rot}=8.43$\,d, 7.95\,d, and 9.47\,d, respectively.}
  \label{period_TESS}
\end{figure*}

%
\begin{table*}
\caption[]{{\em XMM-Newton}/RGS line fluxes of \ihor$^a$}\label{tab:rgsfluxes} 
\tabcolsep 3 pt
\begin{small}
\begin{tabular}{lrccrcl}
\hline \hline
Ion & $\lambda_{\rm model}$ (\AA)& $\log T_{\rm max}$ & $F_{\rm obs}$ & $S/N$ & Ratio & Blends \\
\hline
\ion{Mg}{xii} & 8.4192 & 7.1 & 1.18e-15 & 3.4 & -0.36 & \ion{Mg}{xii}  8.4246 \\
\ion{Mg}{xi} & 9.1687 & 6.9 & 1.16e-14 & 11.9 & 0.24 & \ion{Mg}{xi}  9.2312 \\
\ion{Mg}{xi} & 9.3143 & 6.9 & 2.27e-15 & 5.3 & -0.13 &  \\
\ion{Ne}{x} & 10.2385 & 6.9 & 3.52e-15 & 7.3 & 0.18 & \ion{Ne}{x} 10.2396 \\
\ion{Fe}{xvii} & 10.5040 & 6.9 & 1.13e-15 & 4.3 & -0.32 & \ion{Fe}{xviii} 10.5364, 10.5382, 10.5640 \\
\ion{Fe}{xvii} & 10.6570 & 6.9 & 9.45e-16 & 4.0 & -0.44 & \ion{Fe}{xix} 10.6001, 10.6116, 10.6193, 10.6840, \ion{Ne}{ix} 10.6426 \\
\ion{Ne}{ix} & 11.0010 & 6.7 & 1.32e-15 & 4.9 & -0.17 & \ion{Fe}{xxiii} 10.9810, 11.0190, \ion{Na}{x} 11.0026, \ion{Fe}{xvii} 11.0260 \\
\ion{Fe}{xvii} & 11.1310 & 6.9 & 1.54e-15 & 5.3 & -0.27 &  \\
\ion{Fe}{xvii} & 11.2540 & 6.9 & 3.60e-15 & 8.2 & -0.07 &  \\
\ion{Fe}{xviii} & 11.4230 & 7.0 & 2.55e-15 & 6.9 & 0.01 & \ion{Fe}{xviii} 11.4226, 11.4254, 11.4274, \ion{Fe}{xvii} 11.4383 \\
\ion{Ne}{ix} & 11.5440 & 6.7 & 1.47e-15 & 5.3 & -0.43 & \ion{Fe}{xviii} 11.5270, \ion{Ni}{xix} 11.5390 \\
\ion{Fe}{xxii} & 11.7700 & 7.2 & 2.97e-15 & 7.7 & 0.17 & \ion{Fe}{xxiii} 11.7360, \ion{Ni}{xx} 11.8320, 11.8460 \\
\ion{Ne}{x} & 12.1321 & 6.9 & 1.84e-14 & 19.7 & -0.12 & \ion{Fe}{xvii} 12.1240, \ion{Ne}{x} 12.1375 \\
\ion{Fe}{xxi} & 12.2840 & 7.2 & 1.18e-14 & 15.9 & 0.06 & \ion{Fe}{xvii} 12.2660 \\
\ion{Ni}{xix} & 12.4350 & 7.0 & 5.43e-15 & 10.9 & -0.03 & \ion{Fe}{xxi} 12.3930, \ion{Fe}{xvi} 12.3973, 12.3983 \\
\ion{Fe}{xvi} & 12.5399 & 6.8 & 1.13e-15 & 5.0 & -0.22 & \ion{Fe}{xx} 12.5260, 12.5760, 12.5760, \ion{Fe}{xvii} 12.5391 \\
\ion{Ni}{xix} & 12.6560 & 7.0 & 4.92e-16 & 3.3 & -0.32 & \ion{Fe}{xxi} 12.6490 \\
\ion{Fe}{xvii} & 12.6950 & 6.9 & 8.47e-16 & 4.3 & 0.09 &  \\
\ion{Fe}{xx} & 12.8460 & 7.1 & 4.72e-15 & 10.3 & -0.01 & \ion{Fe}{xxi} 12.8220, \ion{Fe}{xx} 12.8240, 12.8640, \ion{Fe}{xviii} 12.8430 \\
\ion{Fe}{xx} & 12.9120 & 7.1 & 4.22e-15 & 9.6 & 0.10 & \ion{Fe}{xix} 12.9033, 12.9330, 13.0220, \ion{Fe}{xx} 12.9650 \\
\ion{Fe}{xx} & 13.2740 & 7.1 & 9.26e-16 & 4.6 & -0.09 & \ion{Fe}{xix} 13.2261, 13.2933, \ion{Fe}{xxii} 13.2360, \ion{Fe}{xx} 13.2453, 13.2932, \ion{Ni}{xx} 13.2560 \\
\ion{Fe}{xviii} & 13.3230 & 7.0 & 2.24e-15 & 7.2 & -0.16 & \ion{Ni}{xx} 13.3090, \ion{Fe}{xviii} 13.3550, 13.3807, \ion{Fe}{xx} 13.3850 \\
\ion{Ne}{ix} & 13.4473 & 6.7 & 1.61e-14 & 19.6 & 0.04 & \ion{Fe}{xix} 13.4620, 13.4970 \\
\ion{Fe}{xix} & 13.5180 & 7.1 & 1.10e-14 & 16.3 & 0.18 & \ion{Ne}{ix} 13.5531 \\
\ion{Ne}{ix} & 13.6990 & 6.7 & 9.03e-15 & 14.7 & 0.04 & \ion{Fe}{xix} 13.6450, 13.6878, 13.7054 \\
\ion{Fe}{xvii} & 13.8250 & 6.9 & 1.96e-14 & 21.9 & 0.21 & \ion{Ni}{xix} 13.7790, \ion{Fe}{xix} 13.7950, \ion{Fe}{xvii} 13.8920 \\
\ion{Fe}{xviii} & 13.9530 & 7.0 & 2.05e-15 & 7.1 & 0.05 & \ion{Fe}{xix} 13.9263, 13.9330, \ion{Fe}{xx} 13.9620 \\
\ion{Ni}{xix} & 14.0430 & 7.0 & 6.74e-15 & 12.8 & 0.13 & \ion{Fe}{xxi} 14.0080, \ion{Fe}{xix} 14.0340, 14.0388, \ion{Ni}{xix} 14.0770 \\
\ion{Fe}{xviii} & 14.1580 & 7.0 & 3.72e-15 & 9.5 & 0.55 & \ion{Fe}{xix} 14.1272, 14.1429, 14.1487, 14.1496, \ion{Fe}{xviii} 14.1348 \\
\ion{Fe}{xviii} & 14.2080 & 7.0 & 2.65e-14 & 25.5 & -0.01 & \ion{Fe}{xviii} 14.2560 \\
\ion{Fe}{xviii} & 14.3730 & 7.0 & 1.11e-14 & 16.4 & -0.09 & \ion{Fe}{xviii} 14.3430, 14.3990, 14.4250, 14.4555 \\
\ion{Fe}{xviii} & 14.5340 & 7.0 & 9.28e-15 & 15.2 & 0.05 & \ion{Fe}{xviii} 14.5608,14.5710 \\
\ion{Fe}{xix} & 14.6640 & 7.1 & 2.07e-15 & 7.2 & -0.28 & \ion{Fe}{xviii} 14.6160, 14.6566, 14.6887, \ion{O}{viii} 14.6343, 14.6344 \\
\ion{Fe}{xvii} & 15.0140 & 6.9 & 1.15e-13 & 54.1 & 0.20 &  \\
\ion{O}{viii} & 15.1760 & 6.6 & 8.45e-15 & 14.5 & 0.16 & \ion{Fe}{xvi} 15.1629, \ion{O}{viii} 15.1765, \ion{Fe}{xix} 15.1980 \\
\ion{Fe}{xvii} & 15.2610 & 6.9 & 5.04e-14 & 35.7 & 0.40 &  \\
\ion{Fe}{xvii} & 15.4530 & 6.8 & 1.58e-14 & 25.4 & 0.23 & \ion{Fe}{xviii} 15.3539, 15.4940, \ion{Fe}{xvi} 15.4955 \\
\ion{Fe}{xviii} & 15.6250 & 7.0 & 5.41e-15 & 11.8 & -0.12 & \ion{Fe}{xvi} 15.6398 \\
\ion{Fe}{xviii} & 15.7590 & 7.0 & 2.56e-15 & 8.1 & 0.40 & \ion{Fe}{xvi} 15.7389 \\
\ion{Fe}{xviii} & 15.8240 & 7.0 & 3.47e-15 & 9.4 & -0.28 & \ion{Fe}{xviii} 15.8700 \\
\ion{O}{viii} & 16.0055 & 6.6 & 1.56e-14 & 20.1 & -0.07 & \ion{Fe}{xviii} 15.9310, 16.0040, \ion{Fe}{xvii} 15.9956, \ion{O}{viii} 16.0067 \\
\ion{Fe}{xviii} & 16.0710 & 7.0 & 1.80e-14 & 21.6 & 0.03 & \ion{Fe}{xviii} 16.0450, 16.1590, \ion{Fe}{xix} 16.1100 \\
\ion{Fe}{xvii} & 16.3500 & 6.9 & 3.46e-15 & 9.4 & -0.07 & \ion{Fe}{xvii} 16.2285, \ion{Fe}{xix} 16.2830, 16.2857, \ion{Fe}{xviii} 16.3200 \\
\ion{Fe}{xvii} & 16.7800 & 6.8 & 4.97e-14 & 35.9 & 0.12 &  \\
\ion{Fe}{xvii} & 17.0510 & 6.8 & 1.58e-13 & 63.9 & 0.24 & \ion{Fe}{xvii} 17.0960 \\
\ion{O}{vii} & 17.3960 & 6.5 & 2.93e-15 & 8.6 & 0.43 & \ion{Fe}{xvi} 17.4100, 17.4982, \ion{Cr}{xvi} 17.4207, 17.4252 \\
\ion{Fe}{xviii} & 17.6230 & 7.0 & 3.78e-15 & 9.9 & -0.20 &  \\
\ion{O}{vii} & 18.6270 & 6.4 & 2.82e-15 & 8.5 & -0.06 & \ion{Ca}{xviii} 18.6909 \\
\ion{O}{viii} & 18.9671 & 6.6 & 5.94e-14 & 39.1 & 0.01 & \ion{O}{viii} 18.9725 \\
\ion{N}{vii} & 20.9095 & 6.4 & 8.89e-16 & 4.5 & 0.05 & \ion{N}{vii} 20.9106 \\
\ion{Ca}{xvii} & 21.1560 & 6.9 & 1.90e-15 & 6.7 & -0.02 &  \\
\ion{O}{vii} & 21.6015 & 6.4 & 1.93e-14 & 21.1 & 0.07 &  \\
\ion{O}{vii} & 21.8036 & 6.4 & 3.40e-15 & 8.7 & 0.13 &  \\
\ion{O}{vii} & 22.0977 & 6.4 & 1.29e-14 & 17.0 & 0.01 & \ion{Ca}{xvii} 22.1467 \\
\ion{S}{xiv} & 24.2000 & 6.7 & 5.07e-16 & 3.5 & -0.23 & \ion{Ca}{xvi} 24.2214, \ion{Ca}{xv} 24.2344, \ion{Ca}{xiv} 24.2599, \ion{S}{xiv} 24.2850 \\
\ion{N}{vii} & 24.7792 & 6.4 & 7.53e-15 & 13.5 & -0.01 & \ion{Ar}{xv} 24.7366, 24.7400, \ion{N}{vii} 24.7846, \ion{Ar}{xvi} 24.8509 \\
\ion{Ar}{xvi} & 25.6844 & 6.8 & 8.33e-16 & 4.2 & -0.09 & \ion{Ca}{xiv} 25.7299, 25.7330 \\
\ion{Ar}{xii} & 31.3030 & 6.5 & 1.08e-15 & 3.9 & 0.47 & \ion{Fe}{xviii} 31.3157 \\
\ion{S}{xiv} & 33.5490 & 6.6 & 8.56e-16 & 3.3 & 0.22 & \ion{Si}{xi} 33.5301, \ion{Fe}{xviii} 33.5387 \\
\ion{C}{vi} & 33.7342 & 6.2 & 5.42e-15 & 8.4 & 0.02 & \ion{C}{vi} 33.7396 \\
\ion{Fe}{xvii} & 35.6844 & 6.9 & 2.67e-15 & 5.7 & 0.53 & \ion{Ar}{xiii} 35.7285, \ion{Fe}{xvi} 35.7291, \ion{Ca}{xi} 35.7370 \\
\hline
\end{tabular}

{$^a$ Line fluxes (in erg cm$^{-2}$ s$^{-1}$) 
  measured in {\it XMM-Newton}/RGS \ihor\ spectra, and corrected by the ISM absorption. 
  log $T_{\rm max}$ (K) indicates the maximum
  temperature of formation of the line (unweighted by the
  EMD). ``Ratio'' is the log($F_{\mathrm {obs}}$/$F_{\mathrm {pred}}$) 
  of the line. 
  Blends amounting to more than 5\% of the total flux for each line are
  indicated.}
\end{small}
\end{table*}


%
\begin{table}
\caption[]{{\it HST}/STIS line fluxes of \ihor$^a$}\label{tab:stisfluxes} 
\tabcolsep 4 pt
\begin{small}
\begin{tabular}{lrccrcl}
\hline \hline
Ion & $\lambda_{\rm model}$ (\AA)& $\log T_{\rm max}$ & $F_{\rm obs}$ & $S/N$ & Ratio & Blends \\
\hline
\ion{Si}{iii} & 1206.5019 & 4.9 & 1.11e-13 & 9.5 & -0.44 &  \\
\ion{O}{v} & 1218.3440 & 5.5 & 1.37e-14 & 7.4 & -0.01 &  \\
\ion{N}{v} & 1238.8218 & 5.4 & 1.59e-14 & 7.8 & 0.02 &  \\
\ion{N}{v} & 1242.8042 & 5.4 & 6.94e-15 & 5.2 & -0.03 &  \\
\ion{Si}{ii} & 1264.7400 & 4.5 & 8.82e-15 & 5.6 & 0.26 &  \\
\ion{Si}{ii} & 1265.0040 & 4.6 & 2.55e-15 & 5.1 & -0.11 &  \\
\ion{Si}{iii} & 1298.9480 & 4.9 & 3.55e-15 & 7.5 & 0.81 &  \\
\ion{Si}{ii} & 1309.2770 & 4.6 & 3.56e-15 & 5.3 & 0.01 &  \\
\ion{C}{ii} & 1334.5350 & 4.7 & 3.56e-14 & 12.5 & -0.29 &  \\
\ion{C}{ii} & 1335.7100 & 4.7 & 7.38e-14 & 16.6 & 0.22 & \ion{C}{ii} 1335.665 \\
\ion{Si}{iv} & 1393.7552 & 5.0 & 6.18e-14 & 23.6 & 0.23 &  \\
\ion{Si}{iv} & 1402.7704 & 5.0 & 3.18e-14 & 11.3 & 0.24 &  \\
\ion{Si}{ii} & 1526.7090 & 4.5 & 1.02e-14 & 8.0 & 0.35 &  \\
\ion{Si}{ii} & 1533.4320 & 4.5 & 1.62e-14 & 3.8 & 0.26 &  \\
\ion{C}{iv} & 1548.1871 & 5.1 & 1.04e-13 & 13.1 & 0.03 &  \\
\ion{C}{iv} & 1550.7723 & 5.1 & 4.95e-14 & 13.7 & 0.01 &  \\
\ion{Al}{ii} & 1670.7870 & 4.6 & 2.43e-14 & 5.3 & -0.01 &  \\
\hline
\end{tabular}

{$^a$ Line fluxes (in erg cm$^{-2}$ s$^{-1}$) 
  measured in {\it HST}/STIS \ihor\ spectra, and corrected for the ISM absorption. Columns as in Table~\ref{tab:rgsfluxes}}
\end{small}
\end{table}

\subsection{Determination of flare rate for solar-like {\em Kepler} 
  superflare stars}\label{superflares}
Since we could not find any flare events in the $\sim$52\,d spanned by the \textit{TESS} data of \ihor\ (the data actually cover only 48~d),
we wanted to test the hypothesis that \ihor\ is a superflare star.
To this end we carried out a comparison with a sample of \textit{Kepler} 
superflare stars. From the sample of \citet{bal15}
we selected stars with 
$T_{\rm eff}=5800-6200$\,K and $P_{\rm rot}=5-15$\,d that have shown
at least one flare with energy $>10^{34}$\,erg in the Kepler
short-cadence data (\ihor\ \textit{TESS} observations have a 
two-minute cadence). We found seven stars 
that fulfilled these selection criteria. 

The short-cadence  light curves of this \textit{Kepler} flare star
sample were downloaded from the MAST Portal.
For the analysis we
considered simple aperture photometry (SAP) fluxes. Six out of seven
targets were observed in short-cadence mode only for $1/3$ of an
observing quarter. The short-cadence light curves of one
\textit{Kepler} observing quarter are split into three parts (i.e., for $\sim$30\,d). Only KIC 12004971 has short-cadence data for two
full quarters (quarter 15 and 17) and for an additional $1/3$ of the
time in quarter 1.

\begin{figure}
  \centering
  \includegraphics[width=6cm,angle=270]{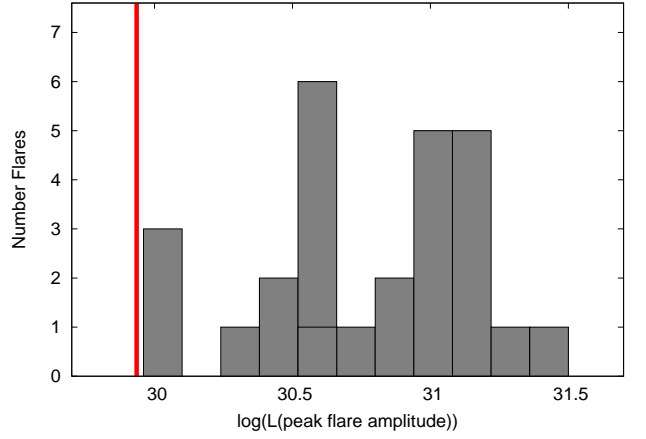}
  \caption{Histogram of the luminosity at the peak emission for the detected
    flares in the \textit{Kepler} superflare star sample. The red
    solid line indicates the flare detection threshold in the {\it
      TESS} observation of \ihor.} 
  \label{fig:kepler_superflares}
\end{figure}

The flare analysis following the procedure outlined in
Sect.~\ref{subsubsect:obs_tess} was performed individually for all
$1/3$ of a quarter. The results are given in Table~\ref{KIC_prop}. In
total we found 27 flares for the seven \textit{Kepler} targets in
12 pieces of $1/3$ quarter. Figure~\ref{fig:kepler_superflares}
shows the histogram of the detected flare amplitudes.
The flare amplitude is measured as the difference between the observed
peak flux and the quiescent flux represented by the light curve
from which rotational signal and outliers were removed \citep[see
  Sect.~\ref{subsect:results_flares} and][]{ste16}.
To show this result in absolute quantities we converted the
relative values of the peak flare amplitude to an amplitude in terms
of luminosity. The {\it TESS} and {\it Kepler} photometries are not flux
calibrated. However, by assuming that the {\it TESS} and {\it
  Kepler} magnitudes 
correspond to the quiescent emission of the star, we could
convert the $T_{\rm mag}$ \citep[from the {\it TESS} candidate target list
  CTL v7.02][]{sta18} and  $K_{\rm p}$ \citep[from the Kepler Input Catalog,
  KIC][]{kep09} to flux using the zero-points and effective
wavelengths provided at the filter profile service of the Spanish
Virtual Observatory \citep[SVO,][]{rod12}. Then we applied the distances given in
Sect.~\ref{sect:intro} for \ihor, and in the CTL for the {\it Kepler}
sample, to 
obtain the quiescent luminosity.  The flare amplitude is then
converted to a luminosity by multiplying the quiescent luminosity
by the relative value of the peak flare amplitude.

The red solid
line in Fig.~\ref{fig:kepler_superflares} indicates the flare
detection threshold for \ihor.
All flares that we found in the {\it Kepler} sample would have been
easily detected in the \ihor\ {\it TESS} light curve.
However, the flare detection threshold for the \textit{Kepler}
stars is on average 2.3 times (0.36 dex in log scale) higher than that
for the \textit{TESS} observation of \ihor.

The individual flare rates of the stars in our \textit{Kepler} sample
range from
$0.03\,{\rm N_{Flares}/day}$ to $0.23\,{\rm N_{Flares}/day}$ with an average of  
$0.08\,{\rm N_{Flares}/day}$.
This latter value is used in Sect.~\ref{subsect:discussion_tess}
to determine the probability of $\iota$\,Hor being a superflare star
despite the absence of such events in the {\em TESS} light curve.

\begin{table}
  \caption{Summary of the properties of our \textit{Kepler} superflare
  star sample}\label{KIC_prop}
  \tabcolsep 4.pt
  \begin{small}
    \begin{tabular}{lcccccc}
      \hline \hline
KIC & $T_{\mathrm{eff}}$ & $P_{\mathrm{rot}}$ & Quarter &
duration & $N_{\mathrm{Flares}}$ & $N_{\mathrm{Flares}}$/day \\
& (K) & (d) & & (d) & & \\
\hline
1025986  & 5966 & 10.076 & 03 & 30.03 & 7 &  0.23 \\
6786176  & 5836 & 7.992  & 04 & 31.05 & 2 &  0.06 \\
7880490  & 6116 & 7.377  & 03 & 30.03 & 6 &  0.20 \\
8556311  & 5886 & 10.222 & 02 & 29.97 & 1 &  0.03 \\
9752973  & 5865 & 13.500 & 03 & 30.34 & 2 &  0.07 \\
12004971 & 5811 & 6.686  & 01 & 33.49 & 1 &  0.03 \\
              &      &        & 15 & 30.83 & 1 &  0.03 \\
              &      &        & 15 & 30.09 & 0 &  0    \\
              &      &        & 15 & 34.91 & 5 &  0.14 \\
              &      &        & 17 & 22.37 & 1 &  0.04 \\
              &      &        & 17 &  4.21 & 0 &  0    \\
12072958 & 5929 & 5.107  & 04 & 31.05 & 1 &  0.03 \\\hline
&&&&& $\varSigma$ 27 & \O{} 0.08 \\
\hline
    \end{tabular}
  \end{small}
  \tablefoot{$T_{\rm eff}$ and $P_{\rm rot}$ are taken from
    \citet{bal15}. Quarter: \textit{Kepler} observing Quarter, where the
    short-cadence data were taken. Duration: time span of short-cadence observations. $N_{\rm Flares}$: number of flares detected in
    the light curve. $N_{\rm Flares}$/day: flare frequency.}
\end{table}

%
%

\end{appendix}
\end{document}